\documentclass[12pt]{article}
\usepackage{amsmath}
\usepackage{amssymb}
\usepackage{latexsym}
\usepackage{epsfig}

\newcommand{\be}{\begin{equation}}
\newcommand{\ee}{\end{equation}}

\newcommand{\sectiono}[1]{\section{#1}\setcounter{equation}{0}}

\newcommand{\LL}{L^\star}

\newcommand{\one}{{\bf [\, \underline 1\,]}}
\newcommand{\two}{{\bf [\, \underline 2\,]}}
\newcommand{\twoa}{{\bf [\, \underline{ 2a}\,]}}
\newcommand{\twob}{{\bf [\, \underline {2b}\,]}}
\newcommand{\twoc}{{\bf [\, \underline {3a}\,]}}
\newcommand{\twod}{{\bf [\, \underline {3b}\,]}}
\newcommand{\three}{{\bf [\, \underline {3}\,]}}

\begin{document}

{}~ \hfill\vbox{
\hbox{CTP-MIT-3755}
\hbox{YITP-SB-06-15}} \break

\vskip 2.0cm
\centerline{\large \bf Solving Open String Field Theory}
\vskip 0.3cm
\centerline{\large\bf  with Special Projectors}

\vspace*{9.0ex}
\centerline{ \sc Leonardo Rastelli}

\vspace*{2.5ex}

\centerline{ \it  C.N. Yang  Institute for Theoretical Physics}
\centerline{ \it Stony Brook University}
\centerline{ \it  Stony Brook, NY 11794, USA}

\vspace*{3.5ex}

\centerline{ \sc Barton Zwiebach}

\vspace*{2.5ex}

\centerline{ \it Center for Theoretical Physics}
\centerline{ \it Massachussetts Institute of Technology}
\centerline{ \it  Cambridge, MA 02139, USA}

\vspace*{6.5ex}
\medskip
\centerline {\bf Abstract }

\bigskip
\bigskip

Schnabl recently found an analytic
expression for the string field tachyon condensate
using a gauge condition
adapted to  the conformal frame
of the sliver projector. We propose that this
construction is more general.
The sliver is an example of 
a {\it special} projector, a projector
such that the Virasoro operator ${\cal L}_0$ and its
BPZ adjoint  ${\cal L}_0^\star$
obey the algebra $[{\cal L}_0, {\cal L}_0^\star] = 
s ({\cal L}_0 + {\cal L}_0^\star)$, with $s$ a positive real constant.
All special projectors provide
abelian subalgebras of string fields,
closed  under both 
the $*$-product and
the action of $\mathcal{L}_0$.
This structure
guarantees exact solvability of a ghost number zero string field equation.
We recast this infinite recursive set of equations
as an ordinary differential equation that is easily solved.  
The  classification of special projectors
is reduced to  a version of the Riemann-Hilbert problem, 
with piecewise constant data on the boundary of a disk.

\bigskip

\vfill \eject

\baselineskip=18pt

\tableofcontents  

\sectiono{Introduction and Summary} \label{s0}

The classical
equation of motion of open string field theory, 
\be \label{eom}
Q_B \Psi + \Psi *  \Psi = 0 \, ,
\ee
is a notoriously complicated system  of infinitely many
coupled equations.
In a recent breakthrough, Schnabl constructed
the first analytic solution of (\ref{eom}), the string field 
that represents the stable vacuum of the open   
string tachyon \cite{Schnabl:2005gv}. 
The solution has been subject to important 
consistency checks~\cite{Okawa:2006vm,Fuchs:2006hw}, it has
been presented in alternative
forms~\cite{Okawa:2006vm},
and it has been used very recently to discuss the absence of physical
states at the tachyon vacuum~\cite{ES}.

In the resurgence of string field theory triggered by
Sen's conjectures on tachyon condensation \cite{Sen:1999mg}, many new techniques
were developed to deal more efficiently with
the open string star product 
\cite{Witten:1985cc}, {\it e.g.} \cite{Rastelli:2000iu}--\cite{Fuchs:2004xj},
see  \cite{Taylor:2003gn, Rastelli:2005mz,Taylor:2006ye} 
for recent reviews.
One line of development \cite{vsft} emphasized
the special role of ``projector'' string fields,
{\it i.e.} string fields squaring to themselves,
\be
\Phi * \Phi = \Phi\,  .
\ee
Particularly simple are   
{\it surface state} projectors \cite{Rastelli:2000iu, Rastelli:2001vb,
Gaiotto:2002kf}.
In the operator formalism of conformal field theory, 
the surface state $\langle f |$
is the state 
associated with the one-punctured disk whose local coordinate
around the puncture is specified by the conformal map $z=f(\xi)$
from a canonical half-disk to the $z$ upper-half plane.\footnote{
The function $z=f(\xi)$ is analytic in the 
 half-disk $H_U = \{   |\xi|  < 1\,, \;\Im \xi \geq 0 \}$, which
is mapped
to a neighborhood of $z=0$, with $f(0)=0$.
The  real boundary of $H_U$ 
   is mapped inside the  boundary of the UHP.}  
If $f(i) =\infty$,
the open string midpoint $\xi=i$ is mapped
to the (conformal) boundary of the disk, and
the corresponding surface state $ \langle f | $ is a projector.
The  sliver \cite{Rastelli:2000iu, Kostelecky:2000hz, Rastelli:2001jb, Rastelli:2001vb} 
and the  butterfly~\cite{Gaiotto:2002kf} 
are the two most studied examples of this construction.
Additional structure is known for the sliver.
There is a continuous family  of wedge states $W_\alpha$, with
$\alpha \geq 0$,\footnote{$ W_\alpha= |\alpha + 1 \rangle$ 
in the notation of \cite{Rastelli:2000iu}.} that
interpolate between the identity string field
$W_0 \equiv {\cal I}$ and the sliver $W_\infty$~\cite{Rastelli:2000iu,Rastelli:2001vb}.
The wedge surface states obey the simple abelian 
multiplication  
rule
\be \label{wedge}
W_\alpha * W_\beta = W_{\alpha + \beta}\, ,  \quad \alpha\, ,
\beta \in [0, \; \infty)\, .
\ee
It seems natural to look
for an analytic solution of (\ref{eom})
that makes use of this abelian family. 
We may look for an expansion of $\Psi$ in terms
of wedge states, with extra ghost insertions necessary 
to obtain ghost number one string fields. However, early  attempts to solve
the equations of motion in the Siegel gauge $b_0 \Psi=0$,
\be
L_0 \Psi+b_0( \Psi * \Psi) = 0 \, , 
\ee
were frustrated by the predicament that while
the star product is  simple in the wedge basis,
the action of $L_0$ is not.\footnote{A closely related difficulty
was encountered in the Moyal formulation of the open string
star product \cite{Bars:2001ag, Douglas:2002jm}. This question has now been
reconsidered in light of the new developments~\cite{Fuchs:2006an}.}
Schnabl's main observation \cite{Schnabl:2005gv} is to choose instead a gauge ${\cal B}_0 \Psi = 0 $
well adapted to the wedge subalgebra.
Here ${\cal B}_0$
is the antighost zero mode in the conformal frame 
$z =f(\xi)= \arctan(\xi)$ of the sliver:
\be
\begin{split}
{\cal B}_0 &\equiv \oint \frac{d z}{2 \pi i}  \,z\, b(z) 
= \oint \frac{d \xi}{2 \pi i} \frac{f(\xi)}{f'(\xi)}
\,b(\xi)  \\[0.5ex]
 &=   \oint \frac{d \xi}{2 \pi i} (1+ \xi^2)
\tan^{-1}(\xi)\, b(\xi) = b_0 + \frac{2}{3}\, b_2 -\frac{2}{15}\, b_4 + \dots
\end{split}
\ee
The corresponding
kinetic operator is the stress tensor zero-mode in the conformal frame of the sliver:
\be
\label{l_zero_sliver}
\begin{split}
{\cal L}_0  &\equiv \oint \frac{d z}{2 \pi i}  \,z\, T(z) 
=  \oint \frac{d \xi}{2 \pi i} \frac{f (\xi)}{f'(\xi)}
T(\xi)  \\[0.5ex]
& =   \oint \frac{d \xi}{2 \pi i} (1+ \xi^2) \tan^{-1}(\xi)\, T(\xi) 
= L_0 + \frac{2}{3}\, L_2 -\frac{2}{15}\, L_4 + \dots \,.
\end{split}
\ee
Here  $T(z)$ is the total stress tensor, which has zero central charge and
is a true conformal primary of  dimension two. 
The operator $\mathcal{L}_0$ has a  simple action on 
the $W_\alpha$ states. In this gauge, the equation of motion
becomes exactly solvable in terms of wedge states (with ghost insertions)!

In this paper we ask how general is the strategy of Schnabl 
 and  investigate its algebraic structure.
What are the algebraic properties of the sliver gauge that ensure
solvability?  Why is a projector relevant?  
For simplicity,
we shall actually focus on the equation of 
motion~\cite{Schnabl:2005gv,Gaiotto:2002uk} 
\be \label{toy}
({\cal L}_0 -1) \Phi + \Phi * \Phi = 0\, ,
\ee
where $\Phi$ has ghost number zero. This
equation captures  many important features of the
full equation of motion~(\ref{eom}). 
We are optimistic
that the algebraic structures discussed in this context
will generalize to (\ref{eom}).
In equation (\ref{toy}), ${\cal L}_0$ is
the zero mode of the stress tensor in some generic
conformal frame $z=f(\xi)$.
We find  an infinite 
class
of local coordinate maps $f(\xi)$ that lead to  solvable equations of the form (\ref{toy}) --
all the maps obeying the three following conditions:

\begin{enumerate}

\item The operator ${\cal L}_0$ and its BPZ conjugate ${\cal L}_0^\star$
satisfy the algebra
\be \label{comm}
[\,{\cal L}_0\,, {\cal L}_0^\star\,] = s\, ({\cal L}_0 + {\cal L}_0^\star)\, ,
\ee
where $s$ is a positive real number.

\item 
The BPZ even operator 
$L^+ \equiv  \frac{1}{s} ( \mathcal{L}_0 + \mathcal{L}_0^\star )$ admits
a non-anomalous left/right decomposition $L^+ = L^+_L + L^+_R$.

\item 
The BPZ odd operators $L^-  \equiv  \frac{1}{s}(\mathcal{L}_0 - \mathcal{L}_0^\star)$
and $K \equiv L^+_L - L^+_R$  annihilate the
identity string field $\mathcal{I}$ of the $*$-algebra.

\end{enumerate}

Condition 1 is a strong restriction on conformal frames.
Conditions 2 and 3, described in detail in section \ref{abelsubalg}, 
can be interpreted as regularity conditions for the vector
$v(\xi)$ associated\footnote{
To a vector field $v(\xi)$ one naturally associates
the linear combinations of Virasoro operators given by the conformally
invariant integral $\int  {d\xi\over 2\pi i} T(\xi) v(\xi)$.}
 with $\mathcal{L}_0$. 
In particular, $v(\xi)$ must vanish at the string midpoint. 

Our analysis of the above conditions for solvability
leads to a nontrivial  
result: the functions $f(\xi)$ that satisfy all of them 
define projectors!
We find this quite interesting, for it explains the relevance of
projectors to solvability.
Not all projectors satisfy all three conditions. 
The projectors
for which conditions 1,2, and 3 are satisfied will be called
{\it special projectors} and will be the focus of this paper.
In summary,  the conformal frames $f$ associated with special projectors
lead to solvable equations.

It is quite easy to find concrete maps that define special projectors. The sliver, of course,
is the original example. 
The operator $\mathcal{L}_0$ in (\ref{l_zero_sliver})  
 satisfies (\ref{comm}) with $s=1$ \cite{Schnabl:2002gg, Schnabl:2005gv}.
But the simplest example is actually the butterfly,
\be
f(\xi) = \frac{\xi}{\sqrt{1+ \xi^2}}  
\quad \to  \quad  {\cal L}_0 = L_0 + L_2 \, , \quad {\cal L}_0^\star = L_0 + L_{-2} \,.
\ee
An elementary calculation reveals that the butterfly
satisfies (\ref{comm}) with $s=2$.

In section 2, we show that for each special projector 
$\langle f|$, there exists an abelian
subalgebra ${\mathcal  A}_f$, closed both under
$*$-product and the action of the kinetic operator ${\cal L}_0$. The subalgebra is
constructed as
\be
{\cal A}_f = {\rm Span} ( \chi_n  ) \, ,
  \quad\chi_n \equiv (L^+)^n \, {\cal I}  \, , \quad n\geq 0 \,.
\ee
The basis states obey
\be \label{chichi}
\chi_m * \chi_n = \chi_{m+n} \,,
\ee
as well as
\be \label{Lchi}
{\cal L}_0 \, \chi_n = s \Bigl( n \chi_n + \frac{1}{2}  \chi_{n+1 }\Bigr)\,.
\ee
Our proof of these properties will be algebraic.
An essential point will be
the decomposition of operators into a
``left'' and ``right''  part, acting on the left and right halves
of the open string.

Interesting elements of ${\cal A}_f$ are the surface states $P_\alpha$,
studied in section~\ref{families_of_interpolating}:
\be
P_\alpha \equiv \sum_{n=0}^\infty \frac{1}{n!} 
\left( -\frac{\alpha}{2} \right)^n \chi_n =
  \exp\left(-\frac{\alpha}{2} L^+ \right) 
\, {\cal I} \,, \quad \alpha \geq 0 \, .
\ee
From (\ref{chichi}),
\be \label{Pab}
P_\alpha * P_\beta = P_{\alpha + \beta} ~.
\ee
Using the sliver's $L^+$, this construction gives the 
familiar wedge states: $P_\alpha \equiv W_\alpha$.
For surface states $\langle f |$ that satisfy conditions 2 and 3  
we have states $P_\alpha$ that satisfy (\ref{Pab}).  The state $P_\infty$,
if it exists, is then a projector.
If $\langle f|$ also satisfies condition 1,  we can prove that
$P_\infty$ coincides with $\langle f|$. Thus we learn that $\langle f|$ is a projector,
in fact, a special projector according to our definition.
In summary, for every special projector $\langle f|$ the $P_\alpha$ 
obey the abelian multiplication rule (\ref{Pab}) and 
interpolate between the identity and $\langle f|$ itself. 
In constructing the above argument we have learned of a general property
of the operator $\mathcal{L}_0$ associated with an arbitrary conformal frame $f(\xi)$: the
corresponding surface state $\langle f|$ can be written  as
\be
\langle f| = \lim_{\gamma\to \infty} \langle \Sigma | \, e^{-\gamma \mathcal{L}_0} \,, 
\ee
where $\langle \Sigma|$ is an arbitrary surface state.  
While any (bra) surface state
is annihilated by its corresponding $\mathcal{L}_0$,
special projectors are also annihilated by $\mathcal{L}_0^\star$.

Properties (\ref{chichi}) and (\ref{Lchi}) guarantee that we
can consistently solve the ghost number zero equation (\ref{toy}) with
$\Phi \in {\cal A}_f$. As described in section 
\ref{solving_equations}, 
this is most efficiently done by considering the ansatz
\be \label{fsans}
\Phi = f_s(x) \, {\cal I}\,,  \qquad x\equiv L^+  \,.
\ee
The equation of motion (\ref{toy}) is translated to  a differential equation for $f_s(x)$,
which is easily solved for all values of $s$. 
For the familiar case of the sliver ($s=1$), Schnabl's solution
of (\ref{toy}) was obtained by recognizing the appearance of 
Bernoulli numbers in an infinite set of recursive equations \cite{Schnabl:2005gv}.  
For us, the solution 
of the $s=1$ differential equation is the generating function of Bernoulli 
numbers.
The solution for $s=2$ is given in terms of  the 
error function and for general
$s$ in terms of hypergeometric functions.  The higher $s$ solutions provide, in
some sense, an $s$-dependent deformation of the Bernoulli 
numbers, which arise from the $s=1$ solution.

Equivalently, in a recursive analysis one can focus on
the Taylor expansion of the function $f_s(x)$ in (\ref{fsans}):
\be \label{fstaylor}
\Phi = f_s(x) \, {\cal I}
= \sum_{n=0}^\infty a_n^{(s)}\,  x^n\,  {\cal I} =  \sum_{n=0}^\infty a_n^{(s)} \chi_n\,.
\ee
We define a  level $\ell^+$ that counts the powers of $L^+$ acting on the
identity. This assigns level~$n$ to the basis states~$\chi_n$. 
The level $\ell^+$ is exactly additive under $*$-product: the product of two
string fields $\Phi_1$ and $\Phi_2$ of definite level satisfies
\be
\ell^+ (\Phi_1 * \Phi_2 ) = \ell^+ (\Phi_1) + \ell^+(\Phi_2) \, .
\ee
 Because of (\ref{Lchi}), 
$\mathcal{L}_0$ acting on a term of definite level gives terms with the
same level and terms with level increased by one.
Thus we can set up  solvable recursion
relations for the coefficients $a^{(s)}_n$:
the term in the equation of motion proportional
to $\chi_N$  depends only on
$a_n^{(s)}$ with $n \leq N$
so that $a_N^{(s)}$ can be determined.
A level is said to be  super(sub)-additive
under a product if the product
of two factors of definite level produces terms with levels greater (less) than or
equal to the sum of levels of the factors. If we have a level
that is super-additive under $*$-product
 and does not decrease by ${\cal L}_0$ action,
we still have an exactly solvable recursion. 
The ansatz in \cite{Schnabl:2005gv},
with basis states  $x^n P_1$ of level $n$, is super-additive under
$*$-product. 
In fact, the product of two such basis states produces states
of arbitrarily higher level. 
Our ansatz (\ref{fsans}) 
leads
to an even simpler recursion, since in our basis the product
is {\it exactly} additive.

The exact solution can also be presented as a linear superposition
of  $P_\alpha$ states acted by $L^+$:
\be
\Phi = P_\infty + \int_0^\infty   d \alpha \, \mu_s(\alpha) \,L^+ \, P_\alpha \, .
\ee
The density function $\mu_s(\alpha)$ turns out to be the inverse Laplace
transform of ${f_s(2x)\over 2x}$.  For $s=1$ we reproduce
the expression obtained in \cite{Schnabl:2005gv}:
$\mu_1(\alpha)$ is a sum of delta functions localized at $\alpha=n$ for 
$n=1,2,3,\ldots \infty$. 
For $s>1$ one has instead a continuous superposition
of $P_\alpha$ states with $\alpha \geq 1$, namely, 
\be
\mu_s(\alpha)=0 \quad  \hbox{for} \quad \alpha <1\,.
\ee
The states $P_\alpha$ with $\alpha <1$ and, in particular, the identity string field $P_0 
= {\cal I}$
are absent in this form of the solution. 
This may be important 
since formal solutions of  string field theory directly based
on the identity ({\it e.g.} \cite{Kluson:2002ex, Kishimoto:2001de,Takahashi:2002ez})
have turned out to be singular.

Having found the exact solution,  we show 
how to re-write it in a better
ordered form,
\be \label{gu}
\Phi = g_s(u)  {\cal I} \, , \quad u = L^\star \,.  
\ee
The function $g_s$ is obtained from $f_s$ as a linear integral transform 
closely related to the Mellin transform. 
 As opposed to $L^+$, the operator $L^\star$ does not contain positively moded
Virasoro operators. From here  it is a short step
to obtain the solution in the ordinary level expansion, that is,
as a linear combination of Virasoro descendants of the SL(2, R) vacuum.

The form (\ref{gu}), expanded in powers of $u$, does not give an
exactly solvable recursion. Indeed, the level $\ell^\star$, defined
by $\ell^\star (u^n {\cal I}) = n$,  is sub-additive under the
$*$-product and the action of $\mathcal{L}_0$ both increases and decreases
the level.  Nevertheless,
a level truncation approximation scheme  is easy to set-up in this basis
because the products take a rather simple form.  In section~5 we show
that $\ell^\star$-level truncation converges very rapidly to the
exact answer. Thus $\ell^\star$-level truncation provides a new 
approximation scheme for string theory.  It is interesting to compare with
the extensively used  
ordinary level truncation \cite{Kostelecky:1989nt, Sen:1999nx, Moeller:2000xv, Gaiotto:2002wy},
 where the level $\ell$ is defined to be 
the eigenvalue of $L_0+1$. 
Ordinary level truncation is less tractable than $\ell^\star$-level truncation.  
While the
kinetic operator in the Siegel gauge preserves ordinary level (by definition!)
the star product of two fields with definite levels generally produces
states of all levels. Moreover, 
the coefficients of the products are complicated to evaluate.

\medskip
All the preceding analysis does not assume any
specific form for the map $f(\xi)$ that defines the special projector.
In section \ref{ex_counter_ex} we look at some concrete
examples of projectors. In \S\ref{fam_proj_nolimit} we
consider a one-parameter family of projectors
and highlight the distinguishing features that
emerge when the parameter is tuned so that
the projector is special. In \S\ref{sec_butt_ex}
we illustrate
 our  general framework 
studying in some detail the example of the butterfly
and some of its generalizations.  We construct explicitly the
family of states $P_\alpha$ that interpolate between the identity
and the butterfly. For large $\alpha$ these states provide
an exact regulator of the butterfly;  previously found regulators
of the butterfly state  only closed approximately under
$*$-multiplication. 

\medskip

In section \ref{hard_work}, we pose the problem of finding the most
general special conformal  frame --  a frame that leads to
the algebra (\ref{comm}), irrespective of whether it also
satisfies conditions 2 and 3. 
 Surprisingly, an analysis that begins
with a second-order differential equation for $f(\xi)$ turns out
to give a linear constraint equation that involves only the values of 
$f(\xi)$ and its complex conjugate on the circle $|\xi|=1$.
This constraint in fact implies that $f(\xi)$ is the analytic
function that solves the  classic Riemann-Hilbert
problem on the interior of a disk for the case of piecewise constant 
data on the circle boundary. It is pleasant
 to find that the question
of the algebra of $\mathcal{L}_0$ and $\mathcal{L}_0^\star$ leads
to such natural mathematical problem.  We do not perform an exhaustive analysis, 
but our results so far suggest that the sliver
is the unique special projector with $s=1$, and that for each integer $s$ there
is only a finite number of special projectors. For $s=2$ we discuss two
special projectors, the butterfly and the moth. Our results suggest a role
for the Virasoro operator $\mathcal{L}_{-s}$ in the 
conformal frame of the projector.
We offer some concluding 
remarks in \S8.
An appendix collects some useful algebraic identities.

\sectiono{The abelian subalgebra}\label{abelsubalg}

\medskip
\noindent
In this section we 
introduce and study the abelian subalgebra
${\cal A}_f$. We begin by defining the operator 
$\mathcal{L}_0$, which can be viewed as the $L_0$ operator
in the conformal frame $z = f(\xi)$:
\be
\label{l0_vect}
 \mathcal{L}_0 = \oint {dz\over 2\pi i}\,  z \,T(z)=
\oint {d\xi\over 2\pi i}\,  {f(\xi)\over f'(\xi)} \,T(\xi)\,.
\ee
The BPZ dual of ${\cal L}_0$ is denoted by 
${\cal L}_0^\star$.\footnote{In~\cite{Schnabl:2005gv} the operator
$\mathcal{L}_0^\star$ is written as the hermitian conjugate $\mathcal{L}_0^\dagger$
of $\mathcal{L}_0$.  We use the $\star$ because, in all generality, the 
algebraic framework requires the use of BPZ conjugation.  For the sliver, BPZ
and hermitian conjugation agree, as they do for all twist even projectors.
If one were to deal with non twist even projectors, one must use
BPZ conjugation.}
A number of formal algebraic properties will be assumed. 
The first property
is that these operators obey the algebra
\begin{eqnarray}
&  \one &  \qquad \qquad   [\,{\cal L}_0\,, {\cal L}_0^\star\,] = 
s ({\cal L}_0 + {\cal L}_0^\star )\,, \quad s> 0 \,.  \nonumber
\end{eqnarray}
We define normalized operators 
\be
L \equiv {1\over s} \mathcal{L}_0 \quad \hbox{and} \quad
\LL \equiv {1\over s} \mathcal{L}_0^\star \,,
\ee
so that the algebra takes the canonical form
\begin{eqnarray}
 \one   \qquad \qquad  [\,L\,, L^\star\,] = L + L^\star\,.  \qquad \qquad \nonumber
\end{eqnarray}
Next, we  form a BPZ even combination $L^+$ and a BPZ
odd combination $L^-$:
\be
L^+ \equiv  L + \LL  \,, \qquad  L^- \equiv  L - \LL \,.
\ee
We shall define a precise notion of the {\it left} and {\it right}
part of an operator, acting on the left and the right half of the open
string. The BPZ even operator $L^+$ is split into a left and a right part
\be
L^+ = L^+_L + L^+_R \,.
\ee
 We demand that formal properties of left/right splitting
actually hold ({\it i.e.} are not anomalous). 
Concretely, we require  
\begin{eqnarray}
\label{twoaaaa}  
&&  \twoa \qquad \qquad [\,L^+_L\,,\, L^+_R\,] = 0 \,, 
\\[0.5ex]
\label{twobbbb}
&& \twob \qquad \qquad L^+_L (\Phi_1 * \Phi_2) =( L^+_L \Phi_1)* \Phi_2  \,.
\end{eqnarray}
Conditions $\twoa$ and $\twob$ describe precisely
condition 2, as stated in the introduction.

Given an operator ${\cal O} = {\cal O}_L + {\cal O}_R$,
we define its {\it dual} $\widetilde {\cal O}$
as
\be
\widetilde {\cal O} \equiv {\cal O}_L - {\cal O}_R \,.
\ee
The duals of $L^+$ and $L^-$ are denoted by $K$ and $J$, respectively, 
\begin{eqnarray}
K \equiv \widetilde {L^+}  & \equiv &   L^+_L - L^+_R \, ,\\
J \equiv \widetilde {L^-}  & \equiv &   L^-_L - L^-_R \,.\nonumber
\end{eqnarray}
Duality, as we will demonstrate, reverses   BPZ parity, so 
$K$ is BPZ odd and $J$ is BPZ even.
The operators $L^-$ and $K$, being BPZ odd, are naively expected to 
be derivations of the $*$-algebra and to annihilate
the identity. As we shall see, these properties can
 be anomalous and must be checked explicitly, so we highlight 
 them as our last formal properties,
\begin{eqnarray}
\label{threeaaaa}
&& \twoc  \qquad \qquad L^- \, {\cal I} =(L - L^\star)\,{\cal I} = 0 \,, 
\\
\label{threebbbb}
&& \twod \qquad \qquad K \, {\cal I} = (L^+_L - L^+_R)\, {\cal I} = 0 \,. 
\end{eqnarray}
Conditions $\twoc$ and $\twod$ describe precisely
condition 3, as stated in the introduction. 

Finally, we define
the subspace ${\cal A}_f$ as
\be \label{Af}
{\cal A}_f = {\rm Span} ( \chi_n  ) \, , \quad
\chi_n \equiv (L^+)^n \, {\cal I}  \, \quad n \geq 0 \,.
\ee
We claim that if all the formal properties hold,
 ${\cal A}_f$ is actually a subalgebra,
closed under both the
$*$-product and  the action of $L$. 
 Let us demonstrate this explicitly:

\begin{itemize}

\item {\it Closure under $L$}. We need to assume only $\one$  and $\twoc$. 
Indeed, 
\be
 L \, \chi_n = L\,( L^+)^n   {\cal I}  = [L,\,( L^+)^n] \, {\cal I}+ (L^+)^n L {\cal I} \,.
\ee
Using $\one$ to compute the commutator 
in the first term and $\twoc$ to re-write the second term, 
\be \label{Lchi2}
L \chi_n = n( L^+)^{n} {\cal I} + \frac{1}{2} (L^+)^{n+1} {\cal I} = n \chi_n + \frac{1}{2} \,\chi_{n+1} \,, 
\ee
as claimed.

\item {\it Closure under $*$}. We need to assume only $\twoa$, $\twob$, and $\twod$.
Indeed, using $\twod$ and $\twoa$ we can write
\be \label{reord}
\chi_n =( L^+)^n \,  {\cal I}  =( L^+)^{n-1} \,(2 L^+_L)  
\, {\cal I}= ( L^+)^{n-k} \,(2 L^+_L )^k  \, {\cal I}= (2 L^+_L )^k\,( L^+)^{n-k}  \,{\cal I} \,
\ee
for any integer $k$, $0 \leq k \leq n$.
Then, from repeated application of $\twob$,
\be
\chi_m * \chi_n = (2 L^+_L)^m | {\cal I}\rangle * \chi_n =(2 L^+_L)^m \, ({\cal I}  * \chi_n) = 
(2 L^+_L)^m \, \chi_n\, ,
\ee
which by (\ref{reord})
 is recognized as  $\chi_{m+n}$. In summary
 \be \label{chichi2}
 \chi_m * \chi_n = \chi_{m+n} \,.
 \ee

\end{itemize}
\medskip  
We can give a simple explanation of the multiplication rule (\ref{chichi2}).
Because of $\twoc$ and $\one$, we see that the
basis states $\chi_n$ are eigenstates of $L^-$:
\be
\textstyle{\frac{1}{2}} L^-\,  \chi_n = \frac{1}{2} [L^-, (L^+)^n ] \,
{\cal I } = n\,  \chi_n \,. 
\ee
The eigenvalue of $\textstyle{\frac{1}{2}} L^-$ will be called the level $\ell^+$, 
so we have $\ell^+ (\chi_n) = n$.
Since $L^-$ is a derivation, it follows that  $\ell^+$ is additive under 
the $*$-product,
thus explaining (\ref{chichi2}).
By contrast, the level in \cite{Schnabl:2005gv} was defined as the eigenvalue
of $L$. The eigenstates of $L$ are $(L^+)^n |P_1 \rangle$
because, as we shall show in \S\ref{Conservation laws}, $L |P_1 \rangle = 0$.
Finally, since $L$ is not a derivation this level is not additive -- it
is in fact super-additive.

\subsection{Vector fields and BPZ conjugation}

To any vector field $v(\xi)$ we associate the stress-energy
insertion ${\bf T} (v)$ defined by
\begin{equation}
\label{vt_def}
{\bf T} (v) \equiv \oint {d\xi\over 2\pi i}\,  v(\xi) \,T(\xi)\,,
\end{equation}
where the integral is performed over some contour that encircles the
origin $\xi=0$. The choice of contour can be important if $v(\xi)$ 
has singularities. 
 In this notation, (\ref{l0_vect}) is written as
\be
\label{tv-lo_not}
\mathcal{L}_0 = {\bf T} (v)\,,  \qquad  v(\xi) = {f(\xi)\over f'(\xi)} \,.
\ee
In this case the vector $v$ does not have singularities for $|\xi|<1$, so the
closed contour can be taken in this domain.
A general identity that follows from the definition (\ref{vt_def})
and the OPE of
two stress tensors (with zero central charge) is
\be
\label{st_comm}
[ \, {\bf T} (v_1)\,, \, {\bf T} (v_2)\,] =  - {\bf T} ( \,[v_1, v_2]\,) \,, \qquad
[v_1, v_2] \equiv  v_1 \partial v_2 - v_2 \partial v_1 \,.
\ee
We will denote by  $({\bf T} (v))^\star$ the BPZ conjugate of ${\bf T} (v)$.
Recall that under BPZ conjugation
$L_n \to  (-)^n L_{-n}$. For
a vector field $v(\xi) = \sum v_n \,\xi^{n+1}$ we find
\be
{\bf T} (v) =   \sum_n v_n L_{n}   \quad \to \quad ({\bf T} (v))^\star
= \sum_n v_n (-1)^nL_{-n} = \sum_n  v_{-n} (-1)^n  L_{n} \,.
\ee
We thus find that
\be
\label{BPZ_tv}
({\bf T} (v))^\star  = {\bf T} (v^\star)\,,
\ee
where the BPZ conjugate 
vector $v^\star$ is
\be
v^\star (\xi) = \sum_n  v_{-n} (-1)^n  \, \xi^{n+1} =
-\,\xi^2\sum_n  v_{n} \,   {1\over (-\xi)^{n+1}}\,.
\ee
We thus recognize that
\begin{equation}
\label{BPZ_vectorp}
v^\star (\xi) = - \xi^2\,  v(-1/\xi)\,.
\end{equation}

Generally, the vector
$v(\xi)$ is analytic in the interior
$|\xi| < 1$ of the unit disk, with possible singularities on the circle
$|\xi|=1$.  It then follows that the BPZ conjugate vector $v^\star$
is analytic outside of the
unit disk with possible singularities on the unit circle.

Vectors of definite BPZ parity must be considered with some care.
For example, given a vector $v$ we can construct the vector  $v^\star$
and then form the vectors
\be
\label{combin_plus_minus}
v^\pm =  v \pm v^\star
\ee
The vector $v^+$ is said to be BPZ even and the vector
$v^-$ is said to be BPZ odd.
The domain of definition of the vectors $v^\pm$
is the common domain of analyticity
of $v$ and $v^\star$.  This domain of $v^\pm$ is the whole
plane for the vector $v=\alpha + \beta \xi + \gamma \xi^2$, with arbitrary constants $\alpha, \beta,$ and $\gamma$.
The domain of $v^\pm$ is an annulus
around the circle $|\xi|=1$ for {\it e.g.} 
 the vector $v = \xi^3$, since $v$
is singular at $\xi=\infty$ while $v^\star$ is singular for $\xi=0$.
When $v$ has branch cuts that emerge from points 
on the circle $|\xi|=1$, we use the  circle (minus the singular
points) as the
domain of $v^\pm$.  
A BPZ even (odd)
vector leads to a BPZ even (odd) operator  ${\bf T}$.

To discuss the symmetry properties of BPZ even/odd operators
we use $t=e^{i\theta}$ for the points on the circle $|\xi|=1$.
Since $1/t=t^*$  (here $*$ is complex conjugation),
we have that (\ref{BPZ_vectorp}) gives
\be
v^\star (t) = -{1\over (t^*)^2} \, v(-t^*) \,.
\ee
It then follows from (\ref{combin_plus_minus}) that
\be
\label{pri_bpz_ref}
{v^\pm (t) \over t} = {v (t) \over t} \pm {v (-t^*) \over (-t^*)}\,.
\ee
Since $t \to -t^*$ is a reflection about the imaginary axis,
we learn that $v^+/t$ is invariant under reflection about the imaginary
axis while $v^-/t$ 
changes sign under this reflection.  If additionally,
$v(-t) = - v(t)$
and $v(t^*) = (v(t))^*$,
then
 \be
\label{pri_bpz_ref_nip}
{v^\pm (t) \over t} = {v (t) \over t} \pm \Bigl({v (t) \over (t)}\Bigr)^*\,.
\ee
In that case 
we see  that $v^+/t$ is real while $v^-/t$ is imaginary.

\subsection{Left/right splitting and duality}

Consider (\ref{vt_def}) 
in which the integral will be performed over the unit circle $C$
defined by $|\xi| = 1$.  For clarity, we continue to use $t$ to represent points
on the $|\xi| = 1$ circle.  We write
\begin{equation}
\label{vt_def_alt}
{\bf T} (v) \equiv \oint_C {dt\over 2\pi i}\,  v(t) \,T(t)\,,
\end{equation}
We call the part of the circle with Re$~t >0$ the left 
part $C_L$  and the part of the circle with Re$~t <0$ the right
part $C_R$.  Associated with a vector $v(t)$ -- in general only
 defined on the unit circle --  we introduce the left part $v_L$
and the right part $v_R$:
\begin{equation}
v_L(t) = 
\begin{cases}
 v(t) \,,  &\text{if $\,t\in C_L$;} \\[0.5ex]
0\,,  &\text{if $\,t \in C_R$;}
\end{cases} \, \quad 
v_R(t) = 
\begin{cases}
0 \,,  &\text{if $\,t\in C_L$;} \\[0.5ex]
v(t)\,,  &\text{if $\,t \in C_R$.}
\end{cases} \, \quad 
\end{equation}  
It is clear from this definition that  
\begin{equation}
v(t ) =  v_L (t) + v_R (t) \,.
\end{equation}
We also write  
\begin{equation}
{\bf T}_L (v) \equiv \int_{C_L} {dt\over 2\pi i}\,  v(t) \,T(t) 
= {\bf T} (v_L)\,,
\qquad  {\bf T}_R (v) \equiv \int_{C_R} {dt\over 2\pi i}\,  v(t) \,T(t)
= {\bf T} (v_R)\,,
\end{equation}
leading to the relation 
\begin{equation}
 {\bf T}_L (v)+ {\bf T}_R (v) = {\bf T} (v)\,.
\end{equation}
Given a vector $v(t)$,
we define the {\it dual} vector field $\widetilde v(t)$ by
reversing the sign of the right part of $v(t)$:
\be
\widetilde v (t) \equiv v_L(t) - v_R(t)\,.
\ee
Analogously, the dual operator $\widetilde{\bf T} ( v)$ is defined as
\begin{equation}
\widetilde{\bf T} ( v) \equiv {\bf T}_L (v) - {\bf T}_R (v) = {\bf T} (\widetilde v)  \,. 
\end{equation}
Note that duality is an involution both for vectors and operators: 
applied twice it produces no change.

We have seen that BPZ even or odd vectors satisfy
even or odd conditions under $t\to - t^*$.  Since this
reflection maps  $C_L$ and $C_R$ into
each other, the BPZ parity of a vector is changed when we change
the sign of the vector over $C_R$ or $C_L$. 
Therefore,  if $v$ has definite BPZ
parity, its dual $\widetilde v$ will have opposite parity,
\be
v = \pm v^\star \longrightarrow  \widetilde v = \mp  (\widetilde v)^\star \,.
\ee

For explicit computations of dual of vector fields we can use the
function $\epsilon (t)$ defined as
\begin{equation}
\epsilon(t) = 
\begin{cases}
 \phantom{-}1 \,,  &\text{if $\,t\in C_L$;} \\[0.5ex]
-1\,,  &\text{if $\,t \in C_R$;}
\end{cases} 
\end{equation}
Multiplication of a vector $v(t)$ by $\epsilon(t)$ changes the sign
of the right part of the vector and thus implements duality.  The function $\epsilon$ has 
the Fourier series
\be
\epsilon(e^{i\theta}) = {2\over \pi}\, \sum_{k\in \mathbb{Z}} 
\,{(-1)^k\over 2k+1}\, e^{i(2k+1)\theta}\,,
\ee
or, equivalently, a Laurent series in $t$:
\be
\epsilon(t) = {2\over \pi}\, \sum_{k\in \mathbb{Z}} 
\,{(-1)^k\over 2k+1}\, t^{(2k+1)}\,.
\ee
We note that 
\begin{equation}
\label{jkxkj}
 \sum_{k\in \mathbb{Z}} 
\,{(-1)^k\over 2k+1}\, t^{2k+1} =  \ldots  +{1\over 5t^5} - {1\over 3 t^3}
 +{1\over t} + t - {t^3\over 3} + {t^5\over 5} + \ldots = 
\tan^{-1} (t) + \tan^{-1} \bigl({1\over t}\bigr)\,. 
\end{equation}
All in all  
\begin{equation}
\label{jdflkj980}
\boxed{\phantom{\Bigg(}\widetilde v (t) = v(t) \epsilon (t) =
v(t) \,\cdot
{2\over \pi} \Bigl[\tan^{-1} (t) + \tan^{-1} \bigl(\,{1\over t}\,\bigr) \Bigr]
\,.  ~~}
\end{equation}
When used to calculate a Virasoro operator, the factor
in brackets must be Laurent expanded.
  
We now specialize this formalism to the operator
 $L^+= L + L^\star$.
Since $L+ \LL$ is BPZ even we can write
\begin{equation}
L^+ \equiv L+ \LL = {\bf T} (v^+) \,, \qquad \hbox{with} \quad ( v^+)^\star (t) = v^+(t)\,. 
\end{equation}
The dual of $L^+$ is the important BPZ odd operator $K$, 
\be
K \equiv \widetilde {L^+} = L^+_L - L^+_R =  {\bf T} (v^+_L) - {\bf T} (v^+_R)
= {\bf T} (\widetilde {v^+}) \,. 
\ee
Since duality is an involution, we also have  
\begin{equation}
\label{need_rel}
L^+ = \widetilde K = K_L  - K_R\,, \quad {\rm with} ~\; K_L = L^+_L \, , ~~K_R = -L^+_R\,.
\end{equation}

\subsection{When do $L^+_L$ and $L^+_R$ commute?}

The function $\epsilon (t)$ 
must be manipulated with some care.  We have, for example
\be
\label{exp_partial_epsilon}
\partial \epsilon =  {2\over \pi}  \sum_{k\in \mathbb{Z}}  (-1)^k t^{2k} = 
{2\over \pi} \Bigl(  \ldots  +{1\over t^4} - {1\over  t^2}
 +1 - {t^2} + {t^4} + \ldots \Bigr)\,.
\ee
The right hand side is almost zero, as is appropriate for the almost
constant function $\epsilon$ whose derivative 
is the sum of two delta functions: 
\be
\label{distribution}
\partial \epsilon = -2 \delta \bigl(\theta - {\pi\over 2}\bigr)  
+ 2\delta \bigl(\theta - {3\pi\over 2}\bigr)\,.
\ee
Note that $\partial \epsilon$ is a BPZ even vector and the associated
stress-tensor is the BPZ even operator
\be
{\bf T} (\partial \epsilon) = {2\over \pi} \sum_{k=0}^\infty (-1)^{k+1} 
( L_{2k+1} - L_{-(2k+1)} )\,.  
\ee
It is also useful to introduce the operator 
associated with $t$ times $\partial \epsilon$:
\be
\label{the_other}
{\bf T} (t\partial \epsilon) = {2\over \pi} \Bigl( L_0
+ \sum_{k=1}^\infty (-1)^{k} 
( L_{2k} + L_{-2k}) \Bigr)\,.  
\ee
One can see from the expansion (\ref{exp_partial_epsilon})
that $t^2 \cdot \partial \epsilon = -\partial \epsilon$.  This is consistent
with (\ref{distribution}) since $t^2= e^{2i\theta} = -1$ for $\theta = \pm \pi/2$.  
It follows that  product of $\partial \epsilon$ and a function
of $t^2$ gives $\epsilon$ times the function
evaluated at $t=i$.\footnote{This property explains why (\ref{the_other}) is
BPZ even. For an arbitrary  vector $v$ of definite BPZ
parity, $t v$ does not have definite BPZ parity.} The product of $t$ times
$\partial \epsilon$, however, is not proportional to $\partial \epsilon$,
thus the necessity for the alternative operator (\ref{the_other}). 
 A  function $\eta (t)$
with a Laurent expansion can be written~as
\be
\eta(t) =  \eta_1(t) + t \eta_2 (t)\,,
\ee  
where both $\eta_1$ and $\eta_2$ contain only even powers of $t$.
It then follows that 
\be
\eta (t ) \partial\epsilon (t) =  \eta_1 (i)\,\cdot \partial \epsilon (t)
+  \eta_2( i ) \cdot  \, t \partial \epsilon (t) \,.
\ee
We say that $\eta(t)$ vanishes {\em strongly} at $t=i$ if both
$\eta_1$ and $\eta_2$ vanish at $t=i$.  If $\eta(t)$ vanishes
strongly at $t=i$, then $\eta(t) \partial \epsilon =0$.
Finally, we note that  $\epsilon(t) \cdot \epsilon (t ) = 1$.  This is
manifest from the definition of $\epsilon$, but can also be checked 
by explicit squaring of the power series (\ref{jkxkj}).

\medskip
We are now ready to discuss 
when  property $\twoa$ holds.  First note that 
\be
[ L^+ , K ]  =  [\,L^+_L + L^+_R\,, L^+_L - L^+_R] =  
- 2 \, [\,L^+_L \,, L^+_R \,] \,.  
\ee
We thus ask, equivalently, when do $L^+$ and $K$ commute?  To answer this,
we compute
\be
[ L^+ , K ] = [\, {\bf T} (v^+) \,, {\bf T} (\, \widetilde{v^+}\,) ]
 = - {\bf T} (\, [ \,v^+, v^+ \epsilon ] ) \,.
\ee
We now note that
\be
[v^+ \,, v^+\epsilon ] =  v^+\partial (v^+\epsilon) - v^+\epsilon \partial v^+ = 
(v^+)^2 \partial \epsilon \,.
\ee
All in all
\be
\label{midpoint_id_one}
\boxed{\phantom{\Biggl(}~
[L^+_L \,, L^+_R \,] =\, {1\over 2} \, {\bf T} ((v^+)^2 \partial \epsilon ) \,.~~}
\ee
The two operators commute if $(v^+)^2$ vanishes
strongly at $t=i$.  The vector fields $v^+$ we will consider are of the form 
$v^+ = t v_2 (t)$, where $v_2$ is a function of $t^2$. 
In this case $L^+_L$ and $L^+_R$ commute when $v^+$ simply vanishes
at $t= i$.

We do not know what additional conditions, if any, are
needed for property $\twob$ to hold. 
The answer, of course,
may depend on the class of states we use in (\ref{twobbbb}).
We leave this question unanswered.

We can readily consider more general commutators.
For example, given two vectors $v$ and $w$ we examine
\begin{equation}
[{\bf T}_L( v)\,, \, {\bf T}_R (w) ]  = [{\bf T}(v_L)\,, \, 
{\bf T} (w_R) ] 
= - {\bf T} \bigl(  \, [ v_L \,,  w_R \,] \bigr)  \,.
\end{equation}
In order to compute the Lie bracket to the right, we note that
\be
v_L =  {1\over 2} (1+ \epsilon)\, v \,, \qquad  w_R =  {1\over 2} (1- \epsilon)\, w\,,
\ee
and using  $\epsilon^2= 1$, 
\be
[v_L \,, w_R] = - {1\over 2} \, v w \, \partial \epsilon\,.
\ee
As a result,
\begin{equation}
\label{midpoint_id_two}
[{\bf T}_L( v)\,, \, {\bf T}_R (w) ] = {1\over 2}
 {\bf T} \bigl(  v w \, \partial \epsilon\bigr)  \,.
\end{equation}
Note that (\ref{midpoint_id_one})  follows from
(\ref{midpoint_id_two}) for when we set both $v$ and $w$ equal to 
$v^+$.  

\bigskip
\noindent
Note now that for vectors $v_1$ and $v_2$  we have
\begin{equation}
[ \widetilde v_1 \,, v_2 ] =  [ v_1 , \, v_2 ]\widetilde{\phantom{A}} - 
v_1 v_2 \, \partial \epsilon \,,
\end{equation}
where the tilde on the right-hand side acts on the full commutator.
It then follows that the corresponding operators ${\bf T}( v_1)$ and 
${\bf T}( v_2)$ satisfy
\be
\label{duality_commutator}
[ \widetilde{\bf T}( v_1),  {\bf T}( v_2)] = 
[ {\bf T}( v_1),  {\bf T}( v_2)]\widetilde{\phantom{A}} 
 + {\bf T} (v_1 v_2 \, \partial \epsilon ) \,. 
 \ee
If we can ignore the midpoint contributions and the algebra
$\one$ holds, we then have
\be
[K, L] = [\widetilde{L^+},  L ] = [L^+ , L]\widetilde{\phantom{A}} =
 - \widetilde{L}^+ = - K  \, .
\ee

\noindent  
{\em Example:}  Given the BPZ odd derivation $K = {\pi\over 2} (L_1 + L_{-1})$ with vector
${\pi\over 2} (1+ t^2)$, the corresponding BPZ even dual vector is
\begin{equation}
\label{xx9lkj}
v^+(t) = (1+ t^2) \cdot \Bigl[\tan^{-1} (t) + \tan^{-1} \bigl(\,{1\over t}
\,\bigr) \Bigr]
\,.  
\end{equation}
We recognize here the vector corresponding to the sliver's $L^+$. 
Since $(v^+)^2 = {\pi^2\over 4} (1+ t^2)^2 $ vanishes for $t=i$, the operators
$L^+_L$ and $L^+_R$ commute (see (\ref{midpoint_id_one})).

\medskip
\noindent
{\em Example:}  For the butterfly
\begin{equation}
L^+ = \frac{1}{2} \,(\mathcal{L}_0 + \mathcal{L}_0^\star ) = 
\frac{1}{2}L_{-2} + L_0 + \frac{1}{2} L_2 =  \widetilde{K}\,,
\end{equation}
for some suitable operator $K$.  Here
\be
v^+(t) =\frac{1}{2}\, t^3 + t + {1\over 2 t}\,.
\ee
Clearly $v^+(t)$ is of the form $t v_2(t)$, with $v_2(t)$ an even function
of $t$ that vanishes for $t=i$.  It follows that $L^+_L$ and $L^+_R$ 
commute. 
The vector dual to $v^+$ is 
\be
\widetilde {v^+} (t) = \Bigl(   \frac{1}{2} t^3 + t + {1\over 2 t} \Bigr) 
{2\over \pi}\, \sum_{k\in \mathbb{Z}} 
\,{(-1)^k\over 2k+1}\, t^{2k+1} \,.
\ee
A short computation gives the simplified form
\begin{equation}
\label{fkk9}
\widetilde {v^+} (t) = -{8\over \pi} \sum_{k\in \mathbb{Z}} {(-1)^k \,\over
(4k^2-1)  (2k+3) } ~ t^{2k+2}\,.
\end{equation}
It follows that
\begin{equation}
\label{fkk}
K ={\bf T}\, (\widetilde {v^+}) =  -{8\over \pi} \sum_{k\in \mathbb{Z}} {(-1)^k \,\over
(4k^2-1)  (2k+3) } ~ L_{2k+1}\,=
 -{8\over \pi} \sum_{k=0}^\infty {(-1)^k \,\over
(4k^2-1)  (2k+3) } ~ K_{2k+1}\,,
\end{equation}
where  $K_n \equiv L_{n} -(-1)^n L_{-n}$ are the familiar
derivations of the $*$-algebra \cite{Witten:1986qs}.
We can then verify explicitly that 
\begin{equation}
[ L^+\,  , K\, ] =  \Bigl[ \, \frac{1}{2}L_{-2} + L_0 + \frac{1}{2} L_2 \, , \,K \,\Bigr]  = 0 \,.
\end{equation}
This confirms
that  
$[L^+_L\, ,\, L^+_R \,] =0$.

\subsection{Derivations and the identity}\label{derivations_and_the_identity}

We now discuss 
 properties $\twoc$ and $\twod$.
The derivations $K_n = L_{n} -(-1)^n L_{-n}$   are well known
to kill the identity 
string field,
\be \label{Kid}
\langle {\cal I} | K_n = 0\,.
\ee
In the framework
of conservation laws discussed in~\cite{Rastelli:2000iu}, 
 for an operator  ${\bf T} (v)$ to
 annihilate the identity 
the vector $v$  must meet two
conditions. First,  it must be BPZ odd,
\be
{v (t) } = \frac{v(-t^*)}{(t^*)^2}\,.
\ee
This condition states that $v$ is consistent with the gluing condition
of the identity.  Second, when referred to the identity
conformal frame $z = 2\xi/(1-\xi^2)$, the vector 
 $ v (z) $ must be an analytic function
 everywhere
except at $z=0$, where it may have poles.
  Both conditions can be  checked for the vector  
 $v_{n} (\xi) = \xi^{n+1} - (-1)^n \xi^{-n+1}$ 
corresponding to $K_n$.
  
Clearly, by the same argument, any finite linear combination of the $K_n$'s 
will also kill the identity. Subtleties may arise when we consider 
infinite linear combinations.  We believe that the analyticity condition
 for $v(z)$ that we just stated  is stronger than needed: BPZ odd vectors 
with mild  singularities 
still define operators that kill the identity. 
A proper understanding, which we will not attempt to provide,
may require  generalizing the framework
 of conservation laws to allow for certain kinds 
 of non-analytic vector fields.

\begin{figure}
\centerline{\hbox{\epsfig{figure=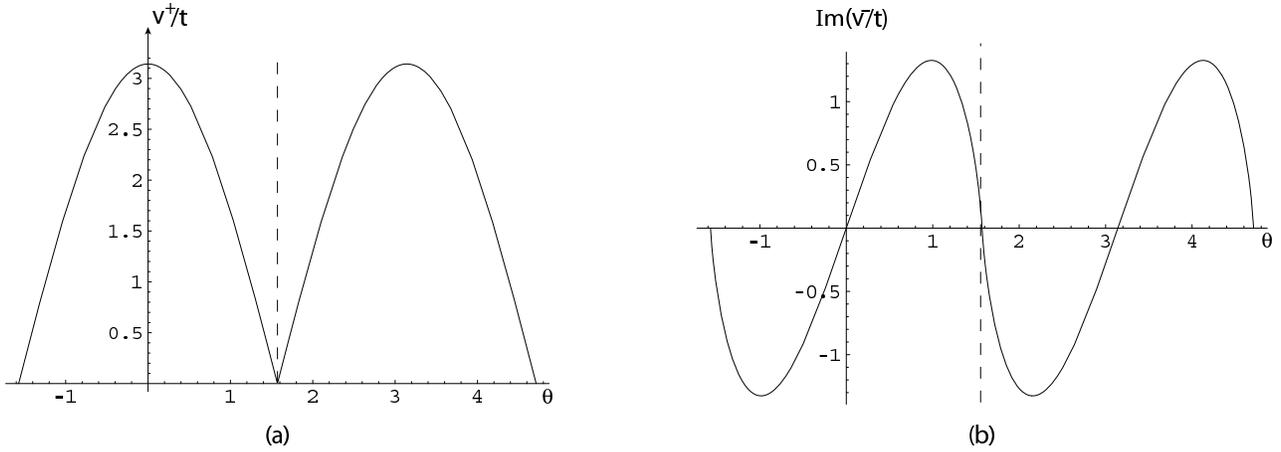, height=5.9cm}}}
\caption{Plot of vectors associated with the sliver's $L^+$ and
$L^-$. (a) Plot of $v^+(t)/t$, with $t= e^{i\theta}$, as a function of $\theta$.
(b) Plot of the imaginary part of $v^-(t)/t$ as a function of $\theta$.  
Both $v^+$ and $v^-$ vanish at the string midpoint $\theta = \pi/2$.}
\label{rz5fig}
\end{figure}

Let us illustrate this point for the  case of the sliver.
We start with the BPZ even vector $v^+$,   
\be v^+ (t)=
(1+ t^2)  \Bigl[\tan^{-1} (t) + \tan^{-1} \bigl(\,{1\over 
t}\,\bigr) \Bigr] =   (1+ t^2)\, \frac{\pi}{2} \epsilon(t) \, .
\ee
Because of the logarithmic branch cuts
at $t = \pm i$, the domain of definition of $v^+$
is only  the unit circle. Figure~\ref{rz5fig}(a)
shows a plot of $v^+ (t)/t$ as a function of $\theta$
($t= e^{i\theta}$). We note that there are
corners  at $\theta=\pm \pi/2$: the derivative fails
to be continuous at this point.  
  Interestingly, 
the singularities are erased in taking
the dual vector $\widetilde {v^+} = \epsilon(t) v^+(t) = \frac{\pi}{2}(1+t^2)$,
where we have used $\epsilon(t)^2 =1$. The
operator $K = \frac{\pi}{2} K_1$ 
certainly kills the identity --
no issue here.  On the other hand, the situation
for $L^-$ is more subtle.
 The corresponding  BPZ odd vector $v^-$ is
 \be v^-(t) = 
(1+ t^2) \Bigl[\tan^{-1} (t) - \tan^{-1} \bigl(\,{1\over \
t}\,\bigr) \Bigr] \,. 
\ee
This vector (plotted in Figure~\ref{rz5fig}(b)) vanishes  
at the string midpoint $t = \pm i$ where it has a first order 
zero multiplied
by a divergent  logarithm.
 The vector is only defined on the unit circle.
 Nevertheless,  
$L^-$, an infinite linear combinations of $K_n$ derivations,
 actually kills the identity. We are going to prove
 this fact in \S 3.2.

For the butterfly the
situation is similar, with the roles
of $K$ and $L^-$ in some sense reversed.
 The operators $L^\pm $ are
 finite  
 linear combinations of Virasoro operators.
 The BPZ odd combination $L^-$ is just
 $K_2$ and certainly kills the identity.  
While the singularity of the $\epsilon$ function  
shows up  in the dual vector $\widetilde {v^+}$ that
 corresponds to $K$, it is quite mild: $\widetilde {v^+}/t$ 
and its derivative vanishes at the midpoint. Thus 
the butterfly $K$ is even less singular
than the sliver $L^-$, and we
 believe that it annihilates the identity.
 
 We will encounter in \S\ref{s=1subsection} 
examples
 of BPZ odd vector fields that, we suspect,  fail to  
  kill the identity. In those cases the midpoint singularity
 is a zero with fractional power.
 By contrast, in all concrete examples of special projectors that we
 are aware of, 
the BPZ odd vector fields have integer power zeroes and, at most,
 logarithmic singularities at the midpoint.

While condition $\twoa$  led to a clear constraint
on the midpoint behavior of the vector $v(\xi)$ associated with $\mathcal{L}_0$, 
we do not    
 know what  constraints on $v(\xi)$ are imposed by  $\twoc$ and $\twod$. It may be that fractional
power zeroes are not allowed for $v(\xi)$ anywhere on the unit circle.

\sectiono{Families of interpolating states}\label{families_of_interpolating}

We now define a family of  states parameterized by a
real constant $\alpha \in [0, \infty)$:
\begin{equation}
\label{palpha_states}
\langle P_\alpha | \equiv \langle \mathcal{I} | 
\, e^{-{\alpha\over 2} (L+ \LL)}  = \langle \mathcal{I} | 
\, e^{-{\alpha\over 2} L^+} \,.
\end{equation}
Since the operator $L^+$ is BPZ even, we also have
\begin{equation}
\label{palpha_states9}
| P_\alpha \rangle =   e^{-{\alpha\over 2} (L+ \LL)}\, |\mathcal{I} \rangle 
 = e^{-{\alpha\over 2} L^+}   | \mathcal{I} \rangle 
\,  \,.
\end{equation}
Unlike the generic elements in the abelian subalgebra $\mathcal{A}_f$,
the states $|P_\alpha \rangle$ have a geometric interpretation as surface
states. 
If the operator $L$ is defined using the conformal frame of the 
sliver the $|P_\alpha\rangle$ states are simply the 
familiar wedge states:
$|n \rangle = |P_{n-1}\rangle$, with $|n=2 \rangle = |P_1\rangle$
equal to the SL(2,R) vacuum and with $|P_\infty\rangle$ the sliver state. 
In general, the family of states
$|P_\alpha\rangle$ interpolates between the identity, for $\alpha=0$,
and the limit state $|P_\infty\rangle$. 
In this section we discuss properties of the above families of states.

\subsection{Star multiplication in the family}

As explained in \S\ref{abelsubalg}, if conditions $\two$ hold we have the abelian algebra
\be \label{abelianagain}
(L^+)^n | {\cal I}\rangle * (L^+)^m | 
{\cal I}\rangle =(L^+)^{m+n} | {\cal I}\rangle \,,
\ee
and we immediately find
\begin{equation}
\label{family_algebra}
|P_\alpha\rangle * |P_\beta\rangle =  |P_{\alpha+ \beta} \rangle \,.
\end{equation}
Notice that for this we do not need to assume the
algebra $\one$. 
We can also prove (\ref{family_algebra}) directly. 
Using  $\twoa$, $\twob$, and $\twod$
we have 
\begin{equation}
|P_\alpha\rangle =
e^{ -{\alpha\over 2} (L^+_L + L^+_R) }
\, |\mathcal{I}\rangle = 
e^{ -{\alpha\over 2} L^+_L }e^{ -{\alpha\over 2} L^+_R }
\, |\mathcal{I}\rangle\,=
e^{ -\alpha\, L^+_L }|\mathcal{I}\rangle =  e^{ -\alpha\, L^+_R }
\, |\mathcal{I}\rangle\, ,
\end{equation}
and therefore
\begin{equation}
|P_\alpha\rangle * |P_\beta\rangle = (e^{ -\alpha\, L^+_L }|\mathcal{I}\rangle) *
|P_\beta\rangle  = e^{ -\alpha\, L^+_L } ~\bigl(\,|\mathcal{I}\rangle *
|P_\beta\rangle \bigr) = 
 e^{ -\alpha\, L^+_L }
|P_\beta\rangle  = |P_{\alpha+ \beta} \rangle \,,
\end{equation}
as we wanted to show.  From here we  obtain a slightly generalized
version of  (\ref{abelianagain}) as follows.
First note that 
\be
\label{L+_as_derivative}
L^+ |P_\alpha\rangle  = - 2 {d\over d\alpha} \,|P_\alpha\rangle\,.
\ee
We then have
\be
\begin{split}
(L^+)^m |P_\alpha\rangle  \, * \, (L^+)^n |P_\beta\rangle 
&=  \Bigl(- 2 {d\over d\alpha}\Bigr)^m |P_\alpha\rangle  *
\Bigl(- 2 {d\over d\beta}\Bigr)^n |P_\beta\rangle  \\[1.0ex]
&=  \Bigl(- 2 {d\over d\alpha}\Bigr)^m \Bigl(- 2 {d\over d\beta}\Bigr)^n 
\bigl[ |P_\alpha\rangle  *
|P_\beta\rangle  \bigr] \\[1.0ex]
&=  \Bigl(- 2 {d\over d\alpha}\Bigr)^m \Bigl(- 2 {d\over d\beta}\Bigr)^n 
\, |P_{\alpha+ \beta} \rangle  = (L^+)^{m+n} \,|P_{\alpha+ \beta} \rangle\,. \\[1.0ex]
\end{split}
\ee
In summary,
\begin{equation}
\label{prodrel}
(L^+)^m |P_\alpha\rangle  \,* \, (L^+)^n |P_\beta\rangle 
= (L^+)^{m+n} \,|P_{\alpha+ \beta} \rangle \, . 
\end{equation}
For  $\alpha=\beta =0$ we recover the expected (\ref{abelianagain}).

\subsection{Conservation laws}\label{Conservation laws}

If we now  assume the algebra $\one$ 
we can derive a useful conservation law for the states $\langle P_\alpha|$,
namely, we find an interesting operator that annihilates the states.

Using $\twoc$, 
 \begin{equation}
0 = \langle \mathcal{I} | L^- e^{-{\alpha\over 2} L^+} 
= \langle P_\alpha| e^{{\alpha\over 2} L^+}  L^-
e^{-{\alpha\over 2} L^+}\,.
\end{equation}
The conjugation of $L^-$ is readily evaluated since $[L^+, L^-] = -2L^+$ and
 $[L^+, [L^+, L^-]]$ vanishes: 
\begin{equation}
0 =  \langle P_\alpha|\, [ \,  L^-- \alpha L^+\,]
=   \langle P_\alpha| [\, (1-\alpha) L  - (1+\alpha) \LL\,] \,.
\end{equation}
This is our conservation law.  Taking BPZ conjugate, it is
equivalently rewritten as 
\begin{equation}
\label{cons_law}
0 =  \bigl[ (1-\alpha) \LL  - (1+\alpha) \,L\bigr]~| P_\alpha\rangle\,.
\end{equation}
For $\alpha=1$ this gives 
\begin{equation}
\label{vac_ser}
L | P_1\rangle = 0 \,.
\end{equation}
For the sliver-based family, the state $|P_1\rangle$ is the SL(2,R) vacuum $|0\rangle$.
Note that the SL(2,R) vacuum cannot belong to any other family of states whose limit
state $|P_\infty\rangle$ is not the sliver.  In fact given two families with different
limit states, their only common state is the identity. 

Using (\ref{L+_as_derivative}) 
and the conservation law  (\ref{cons_law})
one can easily verify that both $L$ and $\LL$ have simple action on $|P_\alpha\rangle$:
\begin{equation}
\label{derf}
\begin{split}
L |P_\alpha\rangle  &=  \,\,\,(\alpha-1)\,\, {d\over d\alpha} \,\,|P_\alpha\rangle\,, \\
\LL |P_\alpha\rangle  &=  -(\alpha+1) {d\over d\alpha} \,|P_\alpha\rangle\,.
\end{split}
\end{equation}

In the above, we have assumed that $L^-$ kills the identity
and derived the conservation law (\ref{cons_law}). We could
also run this logic backwards, taking the conservation
law for some value of $\alpha$ as our axiom 
and {\it deduce} $L^- |{\cal I}\rangle = 0$.
For example, taking $L |P_1 \rangle = 0$ as the starting point,
we have
\be
0 = e^{\frac{1}{2}L^+} L\, | P_1 \rangle = e^{\frac{1}{2}L^+} L \, e^{-\frac{1}{2}L^+}\, | {\cal I} \rangle = \frac{1}{2} L^- \,
|{\cal I} \rangle \,.
\ee
In the case of the sliver, 
the statement that $L  | P_1 \rangle = L | 0 \rangle = 0$ is
obvious. Thus for the sliver we have a simple proof that
 $L^-  |{\cal I } \rangle = 0$.

\subsection{The limit state $|P_\infty\rangle$ }\label{limit_state_99}

\noindent
It is interesting to note that the family of states $|P_\alpha\rangle$ in
(\ref{palpha_states}) satisfies the abelian multiplication rule for {\em any}
projector chosen to build the operators $\mathcal{L}_0$ and $\mathcal{L}_0^\star\,$: 
there was no need to assume $\one$ in order to prove
(\ref{family_algebra}).  This raises an interesting question: What is the limit state
$|P_\infty\rangle$?   In this section we prove that the limit state is
the projector itself when the algebra $\one$ holds.  If  $\one$ does
not hold we do not know what $|P_\infty\rangle$ is, or if it is well
defined.  We have verified  in \S\ref{fam_proj_nolimit} that
for a class of projectors that are not special $|P_\infty\rangle$ is
not the projector used to define the family.

\subsubsection{Surface states and $\mathcal{L}_0$  }

To begin our analysis we first discuss a property that shows
the relevance of $\mathcal{L}_0$ to the construction of surface states. 
We want to prove that for {\em arbitrary} conformal frame $f(\xi)$ (not even
a projector!), we can use the corresponding $\mathcal{L}_0$ to write 
the surface state $\langle f|$ as
\be
\label{lo_for_arb_state}
\langle f| = \lim_{\gamma\to \infty} \langle \Sigma | \, e^{-\gamma \mathcal{L}_0} \,, 
\ee
where $\langle \Sigma|$ is an arbitrary surface state.
In order to establish this fact, we must recall how surface states
are written in terms of exponentials of Virasoro operators.

Recall that given a conformal map $\xi \to f(\xi)$ the operator $U_f$ that
implements (via conjugation) the map takes the form  $U_f= \exp ( {\bf T} ({\rm v}))$
where the vector field ${\rm v}(\xi)$ is related to $f(\xi)$ by the Julia equation
${\rm v}(\xi) \partial_\xi f(\xi) = f ({\rm v}(\xi))$~\cite{LC}.  In this case, the surface state
$\langle f|$ associated with the function $f$ is given by $\langle f|= \langle 0| U_f$.
Given ${\rm v}(\xi)$ the function $f(\xi)$ is constructed as
$f(\xi) =g^{-1} ( 1 + g(\xi))$,
where $g'(\xi)= 1/{\rm v}(\xi)$~\cite{Gaiotto:2002kf}. 
We also recall the composition law  $U_f U_g = U_{f\circ g}$
and note that the scaling $\xi\to f(\xi) = e^b \xi$ is realized by the operator
$e^{b L_0}$. 

We now establish a result that will help us prove (\ref{lo_for_arb_state}) and will
also have further utility.

\noindent{\em Claim:} The operator 
$e^{-\gamma\mathcal{L}_0}$, with $\mathcal{L}_0$ 
defined by $f(\xi)$ in (\ref{l0_vect}), realizes the conformal map 
\be
\label{claim_l0_map}
\xi \to h_\gamma(\xi) = f^{-1} (e^{-\gamma} \, f(\xi) \,) \,\,, \quad
\hbox{or} \quad  h_\gamma = f^{-1} \circ e^{-\gamma} \circ f \,. 
\ee
{\em Proof:} First read the value of the vector $\rm v$ and use
the algorithm described above to determine $h_\gamma$.  Since
 $U_{h_\gamma} =\exp{{\bf T}[v]}=e^{-\gamma \mathcal{L}_0}$ we see that
\be
{\rm v}(\xi) = -\gamma \, {f(\xi)\over f'(\xi)}\,.  
\ee
Then, 
\be
{dg\over d\xi}  = - {1\over \gamma} {f'(\xi)\over f(\xi)}  \quad 
\to \quad g(\xi) = -{1\over \gamma} \ln f(\xi) \,.
\ee
The inverse function to $g$ is
\be
g^{-1} (z) =  f^{-1} ( e^{-\gamma\, z} ) \,,
\ee
so we get
\be
h_\gamma (\xi) = g^{-1} ( 1 + g(\xi)) = f^{-1} \bigl( 
e^{-\gamma} \, e^{\ln f(\xi) }\bigr) =   f^{-1} \bigl(  e^{-\gamma} f(\xi) \bigr)\,.
\ee
This completes the proof of the claim. Note that (\ref{claim_l0_map}) implies
\be
\label{op_map_fdf}
U_{h_\gamma}  = 
U_{f^{-1}} \cdot e^{-\gamma L_0} \cdot U_f\,.
\ee

\smallskip
Let us now return to (\ref{lo_for_arb_state}).  Writing $\langle \Sigma| =
\langle 0 | U_\Sigma$, we have 
\be
\label{last_step_proof}
\begin{split}
\lim_{\gamma\to \infty} \langle \Sigma | \, e^{-\gamma \mathcal{L}_0} 
&= \lim_{\gamma\to \infty} \langle 0 |\,U_\Sigma\cdot U_{f^{-1}} 
\cdot e^{-\gamma L_0} \cdot U_f\\[0.5ex]
&=\lim_{\gamma\to \infty} \langle 0 |(e^{\gamma L_0}\cdot
\,U_\Sigma\cdot  e^{-\gamma L_0})\cdot  (e^{\gamma L_0}\cdot U_{f^{-1}} 
\cdot e^{-\gamma L_0}) \cdot U_f
\end{split}
\ee
For any function $g(\xi) = a_g \,\xi + \mathcal{O}(\xi^2)$ one has
\be
U_{g} =  e^{(\ln a_g) L_0}  \cdot \exp \Bigl(\sum_{n=1}^\infty \gamma_n L_n\Bigr)\,,
\ee
with some constants $\gamma_n$. We now note that 
\be
\lim_{\gamma\to \infty} e^{\gamma L_0} \cdot  
U_{g} \cdot e^{-\gamma L_0}  = e^{(\ln a_g) L_0}  \,.
\ee
Indeed, since 
\be
e^{\gamma\,L_0}\cdot  L_n \cdot e^{-\gamma\, L_0} = 
e^{-n \,\gamma} 
\,L_n \,,
\ee
all positively moded Virasoro operators in the expansions of $U_g$ 
are suppressed in the limit $\gamma \to \infty$. 
Back in (\ref{last_step_proof}) with $\Sigma(\xi) = a_\Sigma \xi + \dots$
and $f^{-1} (\xi) = {1\over a_f} \xi + \dots $, 
\be
\label{last_of}
\lim_{\gamma\to \infty} \langle \Sigma | \, e^{-\gamma \mathcal{L}_0} 
=\langle 0 |\,e^{(\ln a_\Sigma) L_0}\, e^{-(\ln a_{f}) L_0} \cdot U_f
= \langle 0| \, U_f =  \langle f|\,,
\ee
as we wanted to show.  Given this representation, it is
clear that the state $\langle f|$
 is unchanged under the action of $e^{- \eta \mathcal{L}_0}$
for infinitesimal $\eta$.  This implies that $\langle f| \mathcal{L}_0 = 0$.
This is a familiar conservation law: the state $\langle f|$  represents
the vacuum in the $f$ conformal frame and $\mathcal{L}_0$ is the zero-mode
Virasoro operator  in that frame.

\bigskip

\subsubsection{Ordering the states $|P_\alpha\rangle$}

To understand better the  surface states $|P_\alpha\rangle$ 
we have to order the  exponential $ \exp(-{\alpha\over 2} L^+)$
in (\ref{palpha_states}). 
The operator $L^+$ is double sided; it contains both positively and negatively
moded Virasoro operators. 
Surface states, however, are usually written as exponentials of single sided
operators acting on the vacuum. When presented as bras, the exponentials include
Virasoro operators $L_n$ with $n\geq 0$.  Such a presentation is known for the
identity, so we must now trade the operator  $ \exp(-{\alpha\over 2} L^+)$
for an operator that involves only $L$.  This can be done if we  
continue to assume that the operators $L$ and $\LL$ satisfy the algebra $\one$. 

We aim to decompose the Virasoro exponential as
\begin{equation}
\label{jkl}
e^{-{\alpha\over 2} (L+ \LL)} =   x^{{{L}^{}}^-} \, y^L  \,,
\end{equation}
where the constants $x$ and $y$ are determined by the value of $\alpha$.
Since $L^- = L- \LL$  kills the identity, this will imply that
\begin{equation}
\langle P_\alpha | \equiv \langle \mathcal{I} | 
\, \,y^L \,.
\end{equation}
We determine $x$ and $y$ as follows.  Using
(\ref{diffls}) equation (\ref{jkl}) is rewritten as
\begin{equation}
e^{-{\alpha\over 2} (L+ \LL)}=\Bigl( {2\over 1+ x^2}\Bigr)^{\LL} 
\Bigl( {2x^2y\over 1+ x^2}\Bigr)^L  \,.
\end{equation}
Comparing with the last equation in (\ref{simpeid}) we deduce that 
\begin{equation}
x^2 = 1/y \quad\hbox{and}\quad {2y\over 1+ y} = {1\over 1+ {\alpha\over 2}}
 \quad \to \quad   y = {1\over 1+ \alpha} \,.
\end{equation}
We therefore have 
\begin{equation}
\label{palpha_ordered}
\langle P_\alpha | \equiv \langle \mathcal{I} | 
\, \,e^{-\gamma\,L }\,, \quad\hbox{with} \quad  \gamma =  \ln (1+ \alpha)\,.
\end{equation}
This provides a conventional presentation for the surface state $\langle P_\alpha|$.

\subsubsection{Conformal frames and the state $|P_\infty\rangle$}

We now wish to determine the conformal frame $z=f_\alpha (\xi)$ associated
with the surface state $\langle P_\alpha|$.  With this information we will
be able to discuss the limit state $\langle P_\infty|$.

For the identity state we have 
\begin{equation}
\langle \mathcal{I} | = \langle 0| U_{f_I} \,, \qquad
f_I(\xi) = {\xi\over (1-\xi^2)}, 
\end{equation}
so, for the  $\langle P_\alpha |$ states (\ref{palpha_ordered}) we have
\begin{equation}
\langle P_\alpha |  = \langle 0| U_{f_I}  U_{h_\gamma} =   \langle 0| U_{f_I\circ h_\gamma}\,, 
\quad \hbox{with} \quad   U_{h_\gamma} = e^{-\gamma\, L} = 
e^{-{\gamma\over s}\, \mathcal{L}_0}  \,.
\end{equation}
It follows from the claim (\ref{claim_l0_map}) proven earlier that 
\be
\label{sol_function}
\langle P_\alpha | = \langle f_\alpha |\,, \quad  \hbox{with} \quad
f_\alpha = f_I \circ f^{-1} \circ e^{-{\gamma\over s}} \circ f \,,
\ee
or for the corresponding operators,
\be
\label{op_map_alpha}
\langle P_\alpha | = \langle 0 |U_{f_\alpha}\,, \quad  \hbox{with} \quad
U_{f_\alpha}= U_{f_I} \cdot  
U_{f^{-1}} \cdot e^{-{\gamma\over s} L_0} \cdot U_f\,.
\ee

\medskip
In order to examine the large $\alpha$ or, equivalently, the large $\gamma$ limit we 
rewrite (\ref{op_map_alpha}) as
\be
\label{op_map_alpha_9}
U_{f_\alpha}= e^{-{\gamma\over s} L_0} \cdot ( e^{{\gamma\over s} L_0}\cdot U_{f_I} \cdot  
e^{-{\gamma\over s} L_0}) \cdot  ( e^{{\gamma\over s} L_0}\cdot
U_{f^{-1}} \cdot e^{-{\gamma\over s} L_0} )\cdot U_f\,.
\ee
Since $f_I$  goes like $f_I(\xi) = \xi + \mathcal{O}(\xi^2)$
and  $f^{-1}(\xi) = {1\over a_f} \xi + \mathcal{O}(\xi^2)$, we find
\be
\label{op_map_alpha_10}
U_{f_\alpha} \simeq  e^{-\bigl( \ln a_f +{\gamma\over s}
\bigr) L_0}  \cdot U_f\,, \quad \hbox{as} \quad  \alpha \to \infty
\ee
Using (\ref{op_map_alpha}) we finally conclude that 
\be \label{limitstate}
\langle P_\alpha | = \langle 0 |U_{f_\alpha}
\simeq \langle 0 |\, e^{-\bigl( \ln a_f +{\gamma\over s}
\bigr) L_0}  \cdot U_f = \langle 0 |U_f  = \langle f|\,
\quad \hbox{as} \quad  \alpha \to \infty\,.
\ee
This is an important result:  the limit state $\langle P_\infty|$ of the
family is the surface state $\langle f|$ whose frame $f(\xi)$ was used
to define $\mathcal{L}_0$. 
 Let us recapitulate the logic. Starting with a conformal frame $f(\xi)$
obeying conditions $\two$ and $\twod$, we were
able to construct the  family of states $|P_\alpha \rangle$
obeying the abelian algebra (\ref{family_algebra}). 
The  algebra implies that if the limiting state $|P_\infty \rangle$ exists,
it is a projector.
By further assuming condition $\one$, we showed
how to reorder the states $|P_\alpha \rangle$
into well-defined surface states, and further proved
that $|P_\infty \rangle = | f \rangle$. All in all,
we reach the nontrivial conclusion that if  the map
$f(\xi)$ satisfies $\one$, $\two$ and $\twod$, it
 defines a projector, indeed
a special projector in our terminology. The role of
condition $\twoc$ is to guarantee that 
${\cal L}_0$ has a simple action on  $|P_\alpha \rangle$,
which is necessary for solvability.

Finally, note that the limit state representation  (\ref{limitstate})
implies that $\langle f |$ is invariant under the action of $e^{-\eta L^+}$,
with infinitesimal $\eta$.  Thus we see that $ \langle f| L^+=0$.
Since any (bra) 
surface state is annihilated by its $\mathcal{L}_0$, we conclude
that special projectors are also annihilated by $\mathcal{L}_0^\star$.

\sectiono{Solving Equations}\label{solving_equations}

We will consider here the string field equation 
\begin{equation}
\label{calLequ}
({\cal L}_0 - 1) \Phi + \Phi * \Phi = 0 \,.
\end{equation}
The operator $\mathcal{L}_0 = L_0 + \ldots\,$, together with $\mathcal{L}_0^\star$
 will be assumed to satisfy
the algebra $\one$.  As we will see, the solution is sensitive
to the value of the constant $s$. 

Before starting let us make a comment concerning reparameterizations.
Under a midpoint preserving reparameterization (a symmetry of the theory)
the string field transforms as
\begin{equation}
\Phi = e^{-K} \Phi' \,, 
\end{equation}
where the generator $K$ is BPZ odd.  Under this change (\ref{calLequ}) becomes 
\begin{equation}
\label{calLequx}
({\cal L}_0' - 1) \Phi' + \Phi' * \Phi' = 0 \,, \qquad \hbox{with} \qquad 
 {\cal L}_0'  \equiv e^K{\cal L}_0\,e^{-K}\,.
\end{equation}
It then follows that 
\begin{equation}
{{\cal L}_0'}^\star = e^K{\cal L}_0^\star \,e^{-K} \,. 
\end{equation}
Since they are related by similarity, the operators
 ${\cal L}_0, {\cal L}_0^\star$ and 
the operators ${\cal L}_0',{{\cal L}_0'}^\star$ define algebras with
the {\em same} value of $s$.  So solutions of the string field
equation for different values of $s$ are not related by reparameterization.
Solutions for different $\mathcal{L}_0$ operators that have the same
value of $s$ might 
be related by reparameterizations, but we will not
attempt to investigate this here.

In order to consider all values of $s$ simultaneously, we use
$L= \mathcal{L}_0/s$ and write (\ref{calLequ}) as
\begin{equation}
\label{keq}
(s\, L -1)\Phi  + \Phi * \Phi = 0 \,.
\end{equation}
All the $s$ dependence is now in the kinetic term -- recall that the
$L$ and $\LL$ operators satisfy a universal, $s$ independent algebra.

To compare our approach with that of Schnabl
let us first recall how the $s=1$ equation was solved in~\cite{Schnabl:2005gv}.
An ansatz was given of the form 
\be 
\Phi =  \sum_{n=0}^\infty  {f_n \over n!}  \, \Bigl(-{1\over 2}\Bigr)^n\,
 (L^+)^n|0\rangle \,,\qquad  
s=1\,.
\ee 
where $L^+$ is understood to be the sliver's $L^+$,  
and the $f_n$ are constants to be determined. For arbitrary special 
projectors, $|P_1\rangle$  plays the role that $|0\rangle$ 
plays for the 
 sliver (see (\ref{vac_ser})).  Thus, for arbitrary $s$
the above must be replaced by 
\be
\label{repl_bern_series}
\Phi =  \sum_{n=0}^\infty {f_n \over n!}  \, \Bigl(-{1\over 2}\Bigr)^n\,
 (L^+)^n|P_1\rangle \,.
\ee 
Following the steps in~\cite{Schnabl:2005gv} we would obtain the recursion  
\bigskip
\begin{equation}
\label{schnabl_rec}
(s \, n  - 1) f_n =  - \sum_{p+q\leq n}  {n!\over p!\,q!\, (n-p-q)!} 
\, \,f_p \,f_q\,.
\end{equation}
For $s=1$ the coefficients that emerge were recognized as Bernoulli
numbers. For arbitrary $s$, the first few recursions above give 
\begin{equation}
\label{gen_ber!}
\begin{split}
f_0 &= 1\,, \\
f_1 &= -{1\over s+1} 
\,,\\
f_2 &=  {1+ 2s -s^2\over (1+s)^2 (1+ 2s)} \,,\\
f_3 & = -{(s-1) (2s^3-9s^2-6s-1)\over (1+ s)^3 (1+2s) (1+ 3s) }  \,.
\end{split}
\end{equation}
For $s=1$ we recover the Bernoulli numbers
$f_0=1, f_1= -{1\over 2}  \,, f_2 = {1\over 6} , 
f_3=0\,, \ldots $, 
while for $s=2$ we find the unfamiliar sequence
$f_0=1,  f_1= -{1\over 3}  \,, f_2 = {1\over 45} , 
f_3 = {11\over 315}\,,
f_4 = {29\over 4725},$ etc. The $f_n(s)$ certainly
define some deformation of the Bernoulli numbers, but it 
seemed difficult to obtain a full solution using this idea.

For $s=1$, equation (\ref{schnabl_rec}) is a variant of the
Euler relation for Bernoulli numbers, which expresses
higher Bernoulli numbers in terms of products of 
lower ones.  The Euler relation can be quickly derived using
a differential equation satisfied by the generating function
of the Bernoulli numbers\footnote{We wish to thank J.~Goldstone
for explaining this to us.}.  To solve (\ref{keq}) we will
derive differential equations for functions of $L^+$.  For
 $s=1$ the solution  will give directly  the generating 
function of Bernoulli numbers.  For other values of $s$ the
solution will be written in terms of confluent  
hypergeometric functions.

\subsection{Deriving  the differential equation} 

We will write the solution to (\ref{keq}) as an arbitrary function $f_s$ of $L^+$
acting on the identity string field:
\begin{equation}
\label{fset}
\Phi =  f_s(x) |\mathcal{I}\rangle \,, \quad     x \equiv  L^+\,.
\end{equation}
Let us first examine the kinetic term. We re-write (\ref{Lchi2}) as
\begin{equation}
L \,  x^n \, |\mathcal{I}\rangle = \Bigl( n\, x^n  + {1\over 2} x^{n+1}
\Bigr) \, |\mathcal{I}\rangle, 
\end{equation}
so acting on functions of $x$:
\begin{equation}
L  f(x) |\mathcal{I}\rangle = \Bigl( 
x {df\over dx} + {x\over 2} \,f \Bigr)  \, |\mathcal{I}\rangle \,.
\end{equation}
 Let us now examine the quadratic term. We have (see (\ref{chichi2}))
\begin{equation}
x^m |\mathcal{I}\rangle\, * \,x^n |\mathcal{I}\rangle =
 x^{m+ n}\, |\mathcal{I}\rangle\, ,
\end{equation}
so for functions we find
\begin{equation}
f(x) |\mathcal{I}\rangle\, * \,g(x) |\mathcal{I}\rangle = 
f(x) g(x)\, |\mathcal{I}\rangle\,.
\end{equation}
With these results, equation  (\ref{keq}) becomes the
following differential equation for  $f_s(x)$:\footnote{
Had we set $\Phi =  h_s(x) |P_1\rangle$, where
 $|P_1\rangle= e^{-x/2} |\mathcal{I}\rangle$, 
the differential equation would be 
\begin{equation}
\Bigl( s \,x{d\over dx} - 1\Bigr) h_s(x) +  h_s^2(x) \, e^{-{x\over 2}} = 0\,.
\end{equation}}
\begin{equation}
\label{Pthede}
\boxed{\phantom{\Biggl(}\Bigl[ s \,x\Bigl( {d\over dx}+ {1\over 2}\Bigr)
 - 1\Bigr]\, f_s(x) +  f_s^2(x) \, = 0 \,.~}
\end{equation}  
We are interested mostly in
solutions of (\ref{calLequ}) for which 
\be
\label{vac_comp_equal_one}
\Phi = |0\rangle + \hbox{Virasoro descendants} \,.
\ee
Indeed, if $\Phi$ has any component along the SL(2,R) vacuum state $|0\rangle$,
the coefficient of this component must be equal to one.  This happens because
$|0\rangle * |0\rangle = |0\rangle +$ descendants, and, 
neither the star product of  descendants nor the star product of a descendant
and the vacuum contain the vacuum. 

We now claim that $x^n |\mathcal{I}\rangle$, with
$n$ positive does not contain the vacuum $|0\rangle$. First note that, with 
zero central charge, the action of $L^+$ on a descendant gives a descendant.
Second, $L^+|0\rangle$ is a descendant. Since the identity state is the
vacuum plus a descendant, it follows that $L^+|\mathcal{I}\rangle$ is a descendant,
and so is $(L^+)^n |\mathcal{I}\rangle$ for any integer $n\geq 1$. We assume
that $f_s(x)$ has a Taylor expansion around $x=0$, so we can conclude that 
the coefficient of $|0\rangle $ in the solution $f_s(x)|\mathcal{I}\rangle$
 is $f_s(x=0)$. Therefore, (\ref{vac_comp_equal_one}) holds if 
\be
\label{bound_cond}
f_s (x=0) = 1  \,. 
\ee

\subsection{Solving the differential equation}

Let us first consider the case $s=1$, which
applies to the sliver and must reproduce the result of~\cite{Schnabl:2005gv}.
Letting $f_1= x/a(x)$ we get:
\begin{equation}
{da\over dx} = {1\over 2}\,  
a + 1  \quad \to \quad  a(x) =  C\, e^{x\over 2} - 2\,,
\end{equation}
where $C$ is an integration constant.  
The condition (\ref{bound_cond}) requires $C=2$, which gives
\begin{equation}
\label{s=1sol}
\Phi = f_1(x) |\mathcal{I}\rangle =
 {x/ 2 \over e^{x/ 2} - 1} \,|\mathcal{I} \rangle \,.
\end{equation}
To compare with~\cite{Schnabl:2005gv}, we use $|\mathcal{I}\rangle
= e^{x/2} |P_1\rangle$ to write 
\begin{equation}
\label{s=1sol_prime}
\Phi = 
 {(-x/ 2) \over e^{-x/ 2} - 1} \,|P_1\rangle \,.
\end{equation}
The function in front of $|P_1\rangle$ is then recognized as
the generating function for Bernoulli numbers:
\begin{equation}
\label{odf_sevd}
{(-x/ 2) \over e^{-x/ 2} - 1} = \sum_{n=0}^\infty {B_n\over n!} \Bigl(
-{x\over 2}\Bigr)^n\,\quad\to\quad \Phi = \sum_{n=0}^\infty {B_n\over n!} \Bigl(
-{x\over 2}\Bigr)^n  |P_1\rangle\,.
\end{equation}
This is exactly the result in (\ref{repl_bern_series}), with
$f_n = B_n$, as it was found in~\cite{Schnabl:2005gv}.  Note the curious
fact that  the form (\ref{s=1sol}) of the solution based  on the
identity also has an expansion governed by Bernoulli numbers, one that does not
have the additional minus signs of (\ref{odf_sevd}): 
\begin{equation}
\Phi = f_1(x)|\mathcal{I}\rangle
  = \sum_{n=0}^\infty {B_n\over n!} \Bigl(
{x\over 2}\Bigr)^n|\mathcal{I}\rangle\,.
\end{equation}
Choosing $C = 2/\lambda$ one finds 
\begin{equation}
\label{var_ter}
f_1(x) =  { \lambda \,(x/2) \over e^{x/2} - \lambda }\,,
\end{equation}
which, for $\lambda <1$, corresponds to the ``pure-gauge" solutions 
of~\cite{Schnabl:2005gv}
and do not contain a component along the vacuum state $|0\rangle$.
For $\lambda =1$, the solution (\ref{var_ter})
 coincides with  (\ref{s=1sol}).

For arbitrary values of $s$ we can solve (\ref{Pthede}) by  writing
\begin{equation}
\label{okjdeuh}
f_s(x) = {x^{1/s} \over a_s(x)} \,.
\end{equation}
We then obtain the first order ordinary differential equation
\begin{equation}
\label{gleaned-sldjk}
a_s' - {1\over 2}\, a_s  = {1\over s} \, x^{{1\over s} -1} 
\quad \to \quad a_s(x) = a_s(0) + {1\over s}\, e^{x\over 2}\int_0^x\, 
u^{\,{1\over s} -1} e^{-{u\over 2}}\,du\,.
\end{equation}
Letting  $u= xt$ we obtain
\begin{equation}
a_s(x) = a_s(0) +  e^{x\over 2} x^{\,{1\over s}}\int_0^1\, dt\, e^{-{xt
\over 2}}\,{d\over dt} t^{\,{1\over s} }\,.
\end{equation}
To ensure (\ref{vac_comp_equal_one}) 
we demand $f_s(x) \to 1$ as $x\to 0$.
This requires $a_s(x) \to x^{1/s}$ which, in turn, 
requires $a_s(0)=0$.  For $a_s(0)\not= 0$ the solution's leading
behavior is $f_s(x) \sim x^{1/s}$ -- this is supposed to be the ``pure-gauge" solution. 
So, nontrivial solutions are
\begin{equation}
f_s(x)  = \Bigl[\,e^{x\over 2} \int_0^1\, dt\, e^{-{xt\over 2}}\,{d\over dt}\, 
t^{{1\over s}} ~\Bigr]^{-1}\,.
\end{equation}
The integral (which is an incomplete Gamma function) can be readily transformed
into a series by successive integration by parts.  One finds
\begin{equation}
f_s(x)  = \Bigl[ {}_1F_1 \Bigl(1, 1 + {1\over s}, {x\over 2}\Bigr) \Bigr]^{-1}\,,
\end{equation}
where ${}_1F_1$ is the confluent hypergeometric function with series expansion
\begin{equation}
{}_1F_1 (a, b, z) = 1 + {a\over b} z  +  {a (a+1)\over b (b+1)} {z^2\over 2!} + \ldots 
\end{equation}
One can write the solution of the string field equation as
\begin{equation}
\label{final_form}
\boxed{\phantom{\Biggl(}
\Phi_s  = \Bigl[ {}_1F_1 \Bigl(1, 1 + {1\over s}, {x\over 2}\Bigr)\Bigr]^{-1} ~  
| \mathcal{I}\rangle\,.~}
\end{equation}
This is the solution of (\ref{calLequ}) for arbitrary $s>0$.  

Since ${}_1F_1 (1, 2, {x\over 2}) = (e^{x/2}-1)/(x/2)$, we recover
the answer for $s=1$. For other values of $s$ the answer cannot be written
in terms of elementary functions.  For $s=2$ we get
\begin{equation}
\label{sesol}
\begin{split}
f_2(x) = {2\over \sqrt{\pi}}  \, {e^{-x/2} \sqrt{x/2} \over  \hbox{Erf}[\sqrt{x/2}]}
&= 1 - {x\over 3}  +{2x^2 \over 45}
-{2x^3\over 945} -{2x^4\over 14175} + {2x^5\over 93555} + \ldots  \\[0.5ex]
&= \Bigl(1 + {x\over 6}  +{x^2 \over 360}
-{11x^3\over 15120} +{29 x^4\over 1814400} + \ldots \Bigr) e^{-x/2} \,.
\end{split}
\end{equation}
The last form was included since it allows one to read the $s=2$ coefficients
$f_n(s)$ discussed below (\ref{gen_ber!}).
 For arbitrary $s$ a series expansion of the solution gives
\begin{equation}
f_s(x)  =  1 - {s\,x\over 2(1+s)}  +{s^3 x^2 \over
4(1+s)^2 (1+ 2s)} -\ldots   \,\,.
\end{equation}

The limit $s\to \infty$ can be evaluated since the hypergeometric function becomes 
${}_1F_1 (1, 1, x/ 2)  = \exp ( x/ 2) $. The solution becomes
\begin{equation}
f_\infty (x) = e^{-x/2} \,,  \qquad \Phi = e^{-x/2}|\mathcal{I}\rangle
=  | P_1\rangle~\,,  \quad   s\to \infty\,.
\end{equation}
Since $L|P_1\rangle = 0$ exactly, the equation is satisfied
because $|P_1\rangle * |P_1\rangle = |P_2\rangle \sim |P_1\rangle$
in the $s\to \infty$ limit.
 In \S 6.2 we will encounter  an infinite family
of special projectors, with ${\cal L}_0 =L_0 + (-1)^{m+1} L_{2m}$,
$s=2m$. It may appear that as $s \to \infty$, this  sequence
of operators converges the ``Siegel gauge'' operator $L_0$.
This is probably naive -- the operators $L_0$ and
$L_0^\star$ coincide, and thus commute
with each other, whereas in the proposed sequence  the commutator becomes
larger and larger. The solution of the ``Siegel gauge'' ghost
number zero equation was computed in level truncation \cite{Gaiotto:2002uk},
and it is clearly not a surface state -- 
unlike our  large $s$ solution $\Phi_\infty = |P_1 \rangle$.

\subsection{Solution as a superposition of surface states}

The solution based on the sliver can also be written as an infinite
sum of derivatives of wedge states $|P_n\rangle$ plus an extra term
sliver state~\cite{Schnabl:2005gv}. Apart from that subtle extra term,
the solution arises by naive expansion of the denominator of (\ref{s=1sol_prime})
in powers of $e^{-x/2}$:
\be \label{s_1_exp}
\Phi = {1\over 2}\, {x\over 1-e^{-x/2}}\, |P_1\rangle = {1\over 2}
\sum_{n=0}^\infty  x\,e^{-nx/2}|P_1\rangle  \, .
\ee  
This expansion is not legal for $x=0$, where $e^{-x/2}=1$.  
Indeed for $x=0$ the right-hand  side of (\ref{s_1_exp}) vanishes term by term,
while the left hand side  is non-zero. 
 Alternatively, the first expression for $\Phi$ contains the
vacuum state while the second does not.  The second expression can be used if we
add to it the state $|P_\infty\rangle$:
\be
\label{the_inspiration}
\Phi = |P_\infty\rangle +  {1\over 2}
\sum_{n=1}^\infty  L^+|P_n\rangle  \, . 
\ee  
This is a solution that contains the vacuum state with unit coefficient.
It 
can be viewed as the sum of two solutions. The first term
 $|P_\infty\rangle$
is a solution because it is annihilated by $L$ and it is a projector.
The second term is a solution because it inherits this property from the
function that gave rise to it.  The sum of solutions is a solution because
$|P_\infty\rangle * L^+ |P_\alpha\rangle = L^+ |P_{\infty + \alpha}\rangle
= L^+ |P_\infty\rangle = 0$. If desired, $L^+$ can be viewed as a derivative
of the surface state using (\ref{L+_as_derivative}). 
Interestingly, only wedge states $|P_n\rangle$ with
$n\geq 1$ contribute. This means that, in some sense,
the solution receives no contribution from the identity string field, nor from
any wedge state $|P_\alpha\rangle$ with $\alpha <1$.

We now explain how to
write the arbitrary $s$ solution as a superposition of $| P_\alpha \rangle$ states.
We will also show that only states with $\alpha\geq 1$ contribute. Inspired by
the structure of (\ref{the_inspiration}) we write
\be 
\label{Psi9Palpha}
\Phi_s = |P_\infty\rangle + 
\int_0^\infty d\alpha \, \mu_s(\alpha) \, L^+| P_\alpha \rangle \,.
\ee
As in the case of $s=1$, we identify the second term above with the
solution obtained from the differential equation,
\be 
\label{PsiPalpha}
f_s(x) |\mathcal{I}\rangle = 
 \int_0^\infty d\alpha \, \mu_s(\alpha) \,x\,e^{-\frac{\alpha}{2} x} 
| {\cal I} \rangle \, .
\ee
This equation requires
\be
\label{lap_trnas_ow}
{f_s(2x) \over 2x} = \int_0^\infty d\alpha \, \mu_s(\alpha) \,e^{-\alpha x}\,,
\ee
which states that $f_s(2x)/(2x)$ is the Laplace transform 
of the density $\mu_s(\alpha)$.\footnote{In the usual notation for Laplace transforms 
$\mu(\alpha)$ corresponds to the signal 
 $G(t)$, which vanishes for $t<0$, and
$f(2x)/(2x)$ corresponds to the Laplace transform $G(s)$.} Equation (\ref{lap_trnas_ow}) implies a familiar
property of Laplace transforms: if
the right-hand side integral converges for some $x_0$
it converges for all $x$ with Re$\,x >\,$Re$\,x_0$.
This implies that the left-hand side must have an abscissa
of convergence.  In fact, we believe that 
$f_s(2x)/(2x)$  is finite for Re$\,x>0$.\footnote{This happens if 
 ${}_1F_1 \bigl(1, 1 + {1\over s}, {x}\bigr)$
has no zero for Re$\,x>0$.  This is readily checked for $s=1$.}

It follows from (\ref{lap_trnas_ow}) that the 
density $\mu_s(\alpha)$ 
 is  the inverse Laplace
transform of $f_s(2x)/(2x)$:
\be
\label{inv_tkgdkl}
\mu_s(\alpha) = \int_{c-i \infty}^{c+ i \infty}\,
d x \, e^{\alpha x} \, {f_s(2x)\over 2x}  \,.
\ee
The real constant $c$ can  be chosen to be any  number greater than zero. 

For large $x$ with Re$\,x>0$ one has the asymptotic behavior:
\be
\label{asympt_behav}
{f_s(2x)\over 2x} \simeq    {1\over 2\Gamma(1+1/s)} \,\, \frac{e^{-x}}{x^{1 -{1\over s}}}  \,.
\ee
This relation, a textbook property of the confluent hypergeometric
function, can be gleaned from eqns.~(\ref{gleaned-sldjk}) and (\ref{okjdeuh}),
recalling that $a_s(0)=0$.  Since $f_s(2x)/(2x)$ is analytic for Re$\,x >0$, the 
contour integral in (\ref{inv_tkgdkl}) can be deformed into a very large semicircle
over which (\ref{asympt_behav}) applies.   It follows that  for $\alpha <1$ and
$s>1$ the integral over the half-circle
goes to zero as the radius of the circle goes to infinity, so we conclude
that
\be
\mu_s(\alpha) = 0\, \quad \hbox{for} \quad \alpha < 1 \, .
\ee
This is what we wanted to establish.  We can perform the inverse Laplace 
transform of (\ref{asympt_behav}) getting
\be
\mu_s(\alpha) \simeq  {1\over 2}
{\sin {\pi\over s} \over {\pi\over s}}\, 
{1\over (\alpha-1)^{1/s}}\, \Theta(\alpha-1) \,,
\ee
where $\Theta(u)$ is the step function:
 $\Theta(u) =1$ for $u \geq 0$, $\Theta(u) = 0$ for $u <0$. Note that the
density vanishes for $\alpha<1$ and it has an integrable singularity at $\alpha=1$.

The case $s=1$ is a bit special and the corresponding $\mu_1(\alpha)$ can be readily found: 
\be
\mu_1(\alpha) =    \int_{c-i \infty}^{c+ i \infty}\,
d x \, e^{\alpha x} \, {f_1(2x) \over 2x}
 ={1\over 2} \int_{c-i \infty}^{c+ i \infty}\,
d x \, e^{\alpha x}  \sum_{n=1}^\infty   e^{-n x} = {1\over 2}
\sum_{n=1}^\infty \delta  (\alpha-n) \, ,
\ee
which back in (\ref{Psi9Palpha}) reproduces (\ref{the_inspiration}).

\begin{figure}
\centerline{\hbox{\epsfig{figure=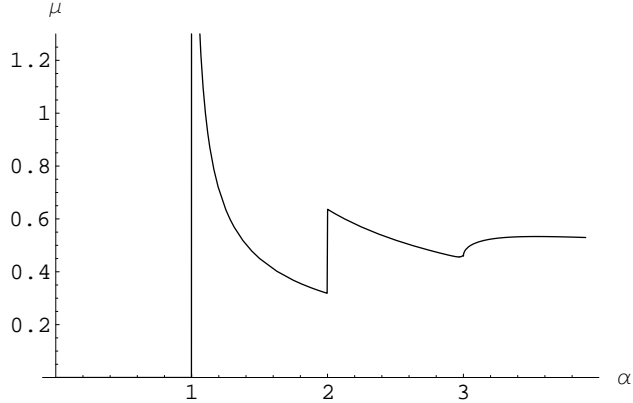, height=5.5cm}}}
\caption{The density $\mu_2(\alpha)$ for the string field solution
of (4.1) with $s=2$. The density vanishes for $\alpha<1$. It has
an integrable singularity for $\alpha=1$, it is discontinuous at
$\alpha=2$ and has a discontinuous derivative at $\alpha=3$. }
\label{rz6fig}
\end{figure}

The density $\mu_2(\alpha)$ for $s=2$ can be 
calculated 
using an  expansion around $x=\infty$.  Using the asymptotic expansion
\be
\hbox{Erf}[\sqrt{x}] = 1 - {e^{-x}\over \sqrt{\pi x}}  \Bigl( 1 + \sum_{n=1}^\infty 
{c_n\over x^n}\Bigr)\,, \quad  
c_n = \bigl( -\textstyle{1\over 2}\bigr)^n \,  (2n-1)!!\,,
\ee
as well as (\ref{sesol}), we can write 
\be
{f_2(2x)\over 2x} = {e^{-x}\over \sqrt{\pi x}}\,\, {1\over \hbox{Erf}[\sqrt{x}]}
= \sum_{m=1}^\infty  \Bigl({e^{-x}\over \sqrt{\pi x}}\Bigr)^{m}
\,\Bigl( 1 + \sum_{n=1}^\infty 
{c_n\over x^n}\Bigr)^{m-1}\,. 
\ee
The inverse Laplace transform can be organized by the exponentials $e^{-(m+1)x}$ each
of which produces a $\Theta (\alpha - (m+1))$.  It follows that $\mu_2(\alpha)$ 
can be calculated 
for $\alpha <4$, for example, by using the  terms with $m=1,2,$ and $3$.
We get
\be
\begin{split}
\mu_2(\alpha) =& ~~ {1\over \pi} { \Theta(\alpha-1)\over \sqrt{\alpha-1}}
+{1\over \pi} \Theta(\alpha-2) \Bigl( 1+ \sum_{n=1}^\infty
{c_n\over n!} (\alpha-2)^n\Bigr) \\
& + \frac{2}{\pi^2}  \,\Theta(\alpha-3) \sqrt{\alpha-3}\, 
 \Bigl(1+ \sum_{n=1}^\infty  \frac{2^n\, d_n}{(2n+1)!!} 
(\alpha-3)^n \Bigr) + \mathcal{O} (\Theta(\alpha-4) ) \,,
\end{split}
\ee
where the coefficients $d_n$ are defined by
\be
d_n = 2 c_n + \sum_{k=1}^{n-1} c_k c_{n-k} \,.
\ee
A plot of the function $\mu_2 (\alpha)$ is shown in Figure~\ref{rz6fig}.\footnote{Since
the series in $\alpha$ that multiplies $\Theta (\alpha-2)$ does not converge
beyond $\alpha=3$, we used (numerical) analytic continuation to construct it in the
range $\alpha\in [3,4]$.}   We believe 
that $\mu_s(\alpha)$ for $s>2$ behaves similarly: it can be built as a sum
of layered step functions, the first of which is multiplied by a function
with an integrable singularity
at $\alpha=1$.

\subsection{Ordering algorithm and descendant expansion}\label{ordering_algo}

In this subsection we show how to use the previously found solutions
to obtain the exact coefficients of a descendant expansion of the 
string field.  Since $L^+$ contains 
both positively moded and negatively moded
Virasoro operators, the solution in the form
\be
\Phi = f(x) | {\cal {I}} \rangle \, ,\quad x =L^+\, ,
\ee
is not suitable for direct evaluation in level expansion.
Indeed, although a convenient level expansion for $|\mathcal{I}\rangle$ is
available \cite{Ellwood:2001ig}
\be
\begin{split}
|\mathcal{I}\rangle  &= 
\ldots 
\exp \Bigl(-{1\over 8} L_{-16}\Bigr) 
 \exp \Bigl(-{1\over 4} L_{-8}\Bigr)   
\exp \Bigl(-{1\over 2} L_{-4}\Bigr) 
\exp (L_{-2}) |0\rangle \,.\\[0.5ex]
&=  |0\rangle + L_{-2} |0\rangle 
+ {1\over 2} L_{-2} L_{-2}|0\rangle  -{1\over 2} L_{-4} |0\rangle  + \ldots \,, 
\end{split}
\ee
the action of functions of $L^+$ on $|\mathcal{I}\rangle$ is complicated.
We need to reorder the expansion by rewriting $\Phi$ as a function of
$\LL$, which does not contain positively moded Virasoro operators (it would 
simplify matters even further if we could remove the $L_0$ from $\LL$).  Thus we
want to use the function $f$ to calculate a function $g$ such that
\be
\Phi =f(x) |\mathcal{I}\rangle =  g(u)  | {\cal {I} }\rangle \,  \quad \hbox{with}
  \quad u \equiv \LL \,.
\ee
We know that $L|\mathcal{I}\rangle = \LL| \mathcal{I}\rangle$.  Therefore, acting on the 
identity, 
\be
x = L^+ =  L+ \LL = 2 \LL  = 2 u \,.
\ee
Since $x u = (u+1) x$ we also have that, acting on the identity,
\be
x^2 =  x (2u) = 2 (u+1) x =  2^2  (u+1) u \,.
\ee
It readily follows from repeated application that, acting on the identity,
\be
\label{xtou}
x^n =2^n (u+n-1) (u+n-2) \ldots (u+1) u  = 2^n \, P_n(u) \,.
\ee
We therefore have the ordering relation (on the identity)
\be
\label{order_relation}
x^n = 2^n \,\frac{\Gamma(u+n)}{\Gamma(u)} \,.
\ee
Note that, in fact, this relation  holds
for $n\geq 0$.  Assuming the Taylor-expansion 
\begin{equation}
f(x)= \sum_{n=0}^\infty f_n x^n\,,
\end{equation} 
and applying the reordering formula (\ref{order_relation})
 to each term of the series, we find 
\be 
\label{R}
g(u) = {1\over \Gamma (u)}
\sum_{n=0}^\infty f_n \; 2^n \, \Gamma (u+ n) 
 = {1\over \Gamma (u)}  \int_0^\infty  dt  \, e^{-t} t^{u-1}
\sum_{n=0}^\infty f_n \,(2t)^n  \,.
\ee
We can thus write $g(u)$ as an integral transform
of $f(x)$ 
\be \label{Rint}
g(u) = \frac{1}{\Gamma(u)}\int_0^\infty dt  
\;  e^{-t }\,t^{u-1}  f(2 t)\, .
\ee
The above is actually the Mellin transform of the function $e^{-t} f(2t)$.  One also
verifies that 
\be
g(0)  =  f(0) \,,
\ee
as expected from the reordering algorithm.

Applying the reordering functional to our solutions (\ref{final_form})
we find 
\be 
\boxed{\phantom{\Biggl(} ~\Phi_s = g_s(u) |\mathcal{I} \rangle =
 \frac{1}{\Gamma(u)}\int_0^\infty dt  \;  
\frac{e^{-t }\, t^{u-1}}{{}_1F_1 \bigl(1, 1 + {1\over s}, {t}\bigr)} ~~|\mathcal{I} 
\rangle \, \,, \qquad  u =\LL \,.~}
\ee
This is our final result for the reordered solution of the string field
equation of motion.  We now consider special values of $s$ to understand
how the result can be used. 
For $s=1$ the integral  becomes
\be 
\Phi_1 = 
 \frac{1}{\Gamma(u)}\int_0^\infty dt  \;  
\frac{e^{-t }\, t^{u}}{e^t -1} ~~|\mathcal{I} 
\rangle \,. 
\ee
This is a familiar integral that can be evaluated in terms of  $\zeta$ functions:
\be 
\label{exact_sol_s=1}
\Phi_1 = 
 u \bigl(\zeta (u+1) - 1\bigr) ~|\mathcal{I} \rangle  \,, \qquad  u = \LL \,.
\ee
This answer is consistent with the form found by Schnabl, whose solution
is written as a function of $\LL$ acting on $|P_1\rangle$. For other finite values
of $s$ one can calculate the function $g_s(u)$ numerically. 
For $s=2$, for example, we obtain
\be
\label{recorded_s2}
g_2(1) = 0.584273\,,\qquad  g_2 (2) = 0.334025\,.
\ee
 As $s\to \infty$ we find $g_\infty (u) = 2^{-u}$ and thus 
\be 
\Phi_\infty =  2^{-u}\,|\mathcal{I} \rangle  \,, \qquad  u = \LL \,.
\ee

\bigskip
The final stage of our calculation requires writing the string field
$g(\LL) |\mathcal{I}\rangle$ as an expansion in Virasoro descendants
-- the familiar level expansion.  We denote  
\be
\label{des_exp}
\Phi= g_s(\LL) |\mathcal{I}\rangle 
 = \gamma_0(s)  |0\rangle + \gamma_2(s) \, L_{-2} |0\rangle 
 + \gamma_4(s) \, L_{-4} |0\rangle 
+ \gamma_{2,2}(s)\, L_{-2} L_{-2} |0\rangle
+\ldots
\ee
with computable $\gamma$ coefficients. 
In here not only is the value of $s$ important but one also requires
the explicit form of the operator  $\LL$. We write
\be
s\LL = L_0 + \alpha_2 L_{-2}  + \alpha_4 L_{-4} + \ldots \,, 
\ee
with constants $\alpha_2, \alpha_4, \ldots$ that take different values for
the different conformal frames.  We now obtain closed-form expressions for
the first few $\gamma$ coefficients in the descendant expansion.
As a first step in the calculation one  shows that for $n\geq 1$
\be
\begin{split}
(s\LL)^n |\mathcal{I}\rangle &=  2^n \, {1\over 2} (\alpha_2 +2) L_{-2} |0\rangle
+ 4^n \, {1\over 4}  (\alpha_4 -2) L_{-4} |0\rangle\\[0.5ex]
& + \Bigl( 4^n \, {1\over 8} (\alpha_2+ 2)^2 - 2^n {1\over 4} \alpha_2 ( \alpha_2 + 2)
\Bigr) \, L_{-2} L_{-2} |0\rangle + \ldots
\end{split}
\ee
One way to obtain the above result is to work out explicitly 
$(s\LL)^n |\mathcal{I}\rangle$
for
$n=1,2,3,$ and $4$ -- the pattern then becomes clear.
Assuming a Taylor expansion  $g(u)= \sum_{n=0}^\infty g_n u^n$ around $u=0$, the
above result leads to (\ref{des_exp}) with
\be
\label{desc_coeff_exact}
\begin{split}
\gamma_0(s) &=  \,\,g_s(0)\,, \\[0.5ex]
\gamma_2(s)&=  -{1\over 2} \,\Bigl[  \,\alpha_2 \,g_s(0) - 
 ( \alpha_2 + 2) \, g_s \Bigl( {2\over s} \Bigr) \Bigr] \,, 
 \\[0.7ex]
\gamma_4(s) &= -{1\over 4}\, \Bigl[  \,\alpha_4 \,g_s(0) - 
 ( \alpha_4 - 2) \, g_s \Bigl( {4\over s} \Bigr) \Bigr] \,,\\[0.7ex] 
\gamma_{2,2}(s)&= \phantom{+}{1\over 8} \,\Bigl[  \,\alpha_2^2 \,g_s(0) - 
 2 \, \alpha_2 ( \alpha_2 + 2) \, g_s \Bigl( {2\over s} \Bigr) 
+ (\alpha_2 + 2)^2 \,  g_s \Bigl( {4\over s} \Bigr) \Bigr] \,. 
\end{split}
\ee
Since the function $g_s(u)$ can be calculated numerically for any $s$ and $u$, the
above descendant expansion can also be obtained numerically. For illustration we
consider the case $s=2$ with the butterfly operator $sL = L_0 + L_{2}$. We thus have
$\alpha_2 =1 $ and $\alpha_4=0$.  Together with $g_2(0)=1$ and the values recorded
in (\ref{recorded_s2}) we get 
\be
\label{des_exp_s2}
\begin{split}
& s=2,~ \mathcal{L}_0 = L_0 + L_2\,:\\
~~&\Phi=   |0\rangle + 0.37641 \, L_{-2} |0\rangle 
 -0.167012\, L_{-4} |0\rangle + 0.062573\, L_{-2} L_{-2} |0\rangle
+\ldots \,.
\end{split}
\ee

\sectiono{$\ell^\star$-level expansion}

Ordinary level truncation is not a very economical
way to solve the string field equations (\ref{calLequ}). 
Indeed, the star multiplication of two Virasoro descendants of
some fixed levels generally gives Virasoro descendants of {\em all} levels,
multiplied by coefficients that take some effort to obtain.
Moreover, convergence to the solutions seems rather slow.
The exact solution obtained in \S\ref{solving_equations} was based on
an expansion in powers of $L^+$ acting on the identity, the power
is then called the level $\ell^+$. The great
advantage of this expansion is that the star product is exactly
additive:  $\ell^+(\Phi_1 * \Phi_2) = \ell^+(\Phi_1) + \ell^+ (\Phi_2)$. 
This allowed us to solve the equation analytically.
The remaining complication was the need to order the solution
when a descendant expansion is needed. The way to order the solution
was explained in \S\ref{ordering_algo}.  

In this section we examine an alternative level truncation.  We 
expand the string field in powers of $\LL$ acting on the identity.
The level $\ell^\star$ of a given term is defined to be the power of $\LL$.  This expansion,
as we will see,
has two advantages over ordinary $L_0$-level expansion in  
Virasoro descendants.  
First, the multiplication
is {\em sub-additive}: $\ell^\star(\Phi_1 * \Phi_2) \leq
 \ell^\star(\Phi_1) + \ell^\star (\Phi_2)$.  
Second, the coefficients appearing in the product are simple
to evaluate.  Moreover, the kinetic term gives no trouble: the $L$ action
on a term of level $l$ gives terms with level less than or equal
to $l+1$.  
Of course, just like for ordinary level truncation, the recursions are
not exactly solvable. 

In the $\LL$ expansion the  string field is written as
\begin{equation}
\label{sflex*}
\Phi = g_s(u) \, |\mathcal{I}\rangle =
 \Bigl(  1 +  a_1 \, u + a_2 u^2  + a_3 u^3 +  \ldots \Bigr) \,
|\mathcal{I}\rangle \,.
\end{equation}
As before, we use the variables
$x =L^+$ and  $u= \LL$ as well as the 
useful relations 
$x\,u = (u+1) \, x$ and $u \, x =  x ( u-1)$
which allow us to move $x$'s and $u$'s across one another.
On the identity $u= x/2$.
In order to evaluate the kinetic term we note that
\begin{equation}
L\, u^n = (x-u) u^n =  x u^n - u^{n+1} =  (u+1)^n (2u) - u^{n+1} \,.
\end{equation}
This implies that for arbitrary functions $g(u)$ we get
\begin{equation}
L \, g(u) |\mathcal{I}\rangle =  u \bigl[  2 g(u+1) - g(u) \bigr] 
\,|\mathcal{I}\rangle\,.
\end{equation}

In order to compute star products we need to invert (\ref{xtou}) 
to find  powers of $u$ expressed in terms of powers of $x$, which
multiply easily. 
For the first few cases we get
\begin{equation}
\begin{split}
 u &= {x\over 2}\,, \\
u^2 &= \Bigl( {x\over 2}\Bigr)^2 -  \Bigl( {x\over 2}\Bigr) \,, \\
u^3 &= \Bigl( {x\over 2}\Bigr)^3-3  \Bigl( {x\over 2}\Bigr)^2  
+\Bigl( {x\over 2}\Bigr)\,,\\
u^4 &= \Bigl( {x\over 2}\Bigr)^4  -6\,\Bigl( {x\over 2}\Bigr)^3 
+ 7 \, \Bigl( {x\over 2}\Bigr)^2  
-\Bigl( {x\over 2}\Bigr)\,.
\end{split}
\end{equation}
In general, we write:
\begin{equation}
u^n =  Q_n \Bigl( {x\over 2}\Bigr) \equiv \sum_{k=1}^n  q_{n,k}  \Bigl( {x\over 2}\Bigr)^k\,, 
\end{equation}
with coefficients $q_{n,k}$ that are defined by
\begin{equation}
q_{\,n,n} = 1 \,,\quad  q_{\,n,1}= (-1)^{n+1} \,, 
\end{equation}
and the recursion relation
\begin{equation}
q_{\,n,k} =  q_{\,n-1, k-1}  \, - \, k \, q_{\,n-1, k}\,,  
\quad  k = 2, \ldots , n-1\,. 
\end{equation}

The star product $u^m \,|\mathcal{I}\rangle  *  u^n \,|\mathcal{I}\rangle$
can be evaluated by  converting {\em one} of the factors explicitly to
the $x$ - basis. Indeed, 
\begin{equation}
u^m \,|\mathcal{I}\rangle  *  u^n \,|\mathcal{I}\rangle
= Q_m \Bigl( {x\over 2}\Bigr)|\mathcal{I}\rangle *
Q_n \Bigl( {x\over 2}\Bigr)\,|\mathcal{I}\rangle  
= Q_n \Bigl( {x\over 2}\Bigr)\,\,Q_m \Bigl( {x\over 2}\Bigr)|\mathcal{I}\rangle
= Q_n \Bigl( {x\over 2}\Bigr)\,\,u^m \,|\mathcal{I}\rangle\,.
\end{equation}
We then have
\begin{equation}
u^m \,|\mathcal{I}\rangle  * u^n \,|\mathcal{I}\rangle = 
\sum_{k=1}^n  q_{\,n,k}  \Bigl( {x\over 2}\Bigr)^k ~ u^m |\mathcal{I}\rangle
= \sum_{k=1}^n  q_{\,n,k} ~(u+k)^m \, P_k (u)  ~  |\mathcal{I}\rangle\,.
\end{equation}
Special cases are
\begin{equation}
\begin{split}
u^m \,|\mathcal{I}\rangle  *  u\,|\mathcal{I}\rangle  
&=  (u+1)^m u \,|\mathcal{I}\rangle\,, \\
u^m \,|\mathcal{I}\rangle  * u^2\,|\mathcal{I}\rangle  
&=  \bigl[(u+2)^m\, u (u+1) - (u+1)^m\, u \bigr]  \,|\mathcal{I}\rangle\,.
\end{split}
\end{equation}

To check the accuracy of the $\LL$ expansion, we examined the cases $s=1$ and
$s=2$. For $s=1$ we obtained,  at various levels,
\begin{equation}
\label{s=1_lev_exp}
\begin{split}
\ell^\star=1: \quad g_1(u) &=  1 - 0.381966 \, u\,, \\
\ell^\star=2: \quad g_1(u) &= 1 - 0.422536 u + 0.0658568 \,u^2\,, \\
\ell^\star=3: \quad g_1(u) &= 1 - 0.422745\, u + 0.0728081 \, u^2 - 0.00518521 \, u^3 \,,\\
\ell^\star=4: \quad g_1 (u) & = 1- 0.422788\, u + 0.0727478 \, u^2 -0.00491614 u^3 
-0.000148237 \, u^4 \,,\\
\ell^\star=5: \quad g_1 (u) & = 1- 0.422788\, u + 0.0728092 \, u^2 -0.00483623 u^3 
-0.000328167\, u^4 \,, \\
\ell^\star=6: \quad g_1 (u) & = 1- 0.422785\, u + 0.0728153 \, u^2 -0.00484579 u^3 
-0.000342449\, u^4 \,. 
\end{split}
\end{equation}
(For levels 5 and 6 we have only written the coefficients up to $u^4$).
We can compare the above with the exact solution 
$g_1(u)$ in (\ref{exact_sol_s=1}), 
that expanded in powers of $u$ gives 
\begin{equation}
g_1(u) =   1-0.422784 \, u  +0.072816\,  u^2-  0.004845\, u^3 - 0.000342 \, u^4 + \ldots\,.
\end{equation}
We see that the convergence of $\LL$ expansion is excellent. The level six results
are very accurate.  We can also examine the coefficient $\gamma_2(1)$ in the 
descendant expansion (\ref{des_exp}) in the sliver conformal frame 
($\alpha_2 = 2/3$).  First note that its exact value is, from (\ref{desc_coeff_exact})
and (\ref{exact_sol_s=1}), 
\begin{equation}
\gamma_2 (1) = -{1\over 3} + {4\over 3} g_1(2) =  {1\over 3} ( 8 \zeta (3) - 9 ) 
\simeq 0.205485  \,.
\end{equation}
The values of $\gamma_2(1)$ obtained from the level expansions in (\ref{s=1_lev_exp})
are found by evaluation of $g_1(2)$. From $\ell^\star=1$ to $\ell^\star=6$ we find
\be
\LL \,- \hbox{expansion}:~~
 \gamma_2 (1) =  -0.018576,~ 0.224474,  ~ 0.20568,  ~ 0.204953, ~0.205375, ~ 0.205428
\,. 
\ee
The convergence, again, is quite good. Had we truncated 
the exact solution, we would have instead obtained the values 
\be
 \hbox{Truncated soln.}:~~
 \gamma_2 (1)= -0.127425,~  0.260926, ~ 0.209244, ~  0.201942, ~ 0.206076,~ 0.205512\,.
\ee
Curiously,  level expansion gets better partial results than the truncated expansion
of the exact solution.

For $s=2$  we get the following results
\begin{equation}
\begin{split}
\ell^\star=1: \quad g_2(u) &=  1 - 0.438447\, u\,, \\
\ell^\star=2: \quad g_2(u) &= 1 - 0.521156 u - 0.102358 \,u^2\,, \\
\ell^\star=3: \quad g_2(u) &= 1 - 0.525210\, u + 0.123726 \, u^2 - 0.0143123 \, u^3\,, \\
\ell^\star=4: \quad g_2 (u) & = 1- 0.524993\, u + 0.124554 \, u^2 -0.016306 u^3 
+0.00101185 \, u^4 \,,\\
\ell^\star=5: \quad g_2 (u) & = 1- 0.524998\, u + 0.124563 \, u^2 -0.0162718 u^3 
+0.000954709 \, u^4 \,. \\
\end{split}
\end{equation}
We clearly seem to have convergence.  
We can calculate the function $g_2$ at $u=1$ and $u=2$ and compare with 
(\ref{recorded_s2}).  Using the level five solution we obtain
\be
\label{recd_s2}
\LL -  \hbox{expansion}: \quad g_2(1) = 0.584248\,,\qquad  g_2 (2) = 0.333357\,,
\ee
in rather good agreement with (\ref{recorded_s2}). And then, for $\ell^\star=1$ up to $\ell^\star=5$ we find
$$ \gamma_2 (2) =  0.342329,~ 0.371804,~ 0.376304, ~0.376401,~0.376402 \,,$$
in nice agreement with the value recorded in (\ref{des_exp_s2}).

We can even work in the limit $s\to \infty$. To order $u^4$, the level ten solution
gives 
\be
g_\infty (u)  = 1- 0.693147 \, u + 0.240226 \, u^2 -0.0555041 u^3 
+0.00961812 \, u^4  + \mathcal{O}(u^5)\,. 
\ee
The expansion of the exact solution is
\be
\begin{split}
g_\infty (u) &= 2^{-u} =  1 - (\ln 2)  \, x + {1\over 2} (\ln 2)^2 \,u^2
 - {1\over 6}(\ln 2)^3\, u^3 + {1\over 24} (\ln 2)^4 \, u^4 
+ \mathcal{O}(u^5)\\[0.6ex]
&\simeq 1-0.693147 \, u + 0.240227  \, u^2 - 0.0555041  \, u^3 + 
0.00961813\, u^4 + \mathcal{O}(u^5)\,.
\end{split}
\end{equation}
The agreement with the level-expansion solution is essentially perfect.

\sectiono{Examples and counterexamples}\label{ex_counter_ex}

In this section we use examples to develop some intuition about
special projectors.  In \S\ref{fam_proj_nolimit}
we begin with a parameterized family of projectors which contains, for special
values of the parameter, three special projectors.  We demonstrate that
only for the special projectors the state $|P_\infty\rangle$  coincides
with the projector. All special projectors are annihilated by the
derivation $K= \widetilde L^+$, since $K$ and $L^+$ commute and $K$ kills
the identity.  We provide  an example in which we demonstrate
that a projector that is not special fails to be annihilated by
$K$.  

 In \S\ref{sec_butt_ex}  we discuss in detail the butterfly
special projector.  We give the explicit form for the frames
$f_\alpha (\xi)$ that define the butterfly family $|P_\alpha\rangle$.
We also discuss regularized butterflies.  We briefly examine
the special projectors that arise from 
$\mathcal{L}_0=  L_0 + (-1)^{m+1} L_{2m}$, $m\geq 1$.

\subsection{A family of projectors}\label{fam_proj_nolimit}

Consider the operator
\be
\label{test_fam}
\mathcal{L}_0=  L_0 + a L_2  + (a-1) L_4\,,
\ee
where $a$ is a real constant whose possible values will
define a family of surface states.  The vector field associated
with $\mathcal{L}_0$ is 
\be
v(\xi, a) =  \xi  + a\xi^3  + (a-1) \xi^5\,.
\ee   
The coefficients have been adjusted so that $v(i, a)=0$ -- the 
vector field vanishes at the string midpoint.  The conformal
frame $z=f(\xi,a)$ can be obtained by solving the differential
equation
\be \label{a_family}
\partial_\xi \ln f(\xi, a) = {1\over v(\xi, a)} \quad \to \quad
f(\xi, a) = {\xi\over \bigl[ 1+\xi^2\bigr]^{1\over 2(2-a)}} 
~ \Bigl(1+ (a-1) \xi^2 \Bigr)^{a-1\over 2(2-a)}\,.
\ee
We require the $f(\xi,a)$ to have no singularities for $|\xi|<1$.  This
fixes $0 \leq  a \leq 2$, but since the exponents become infinite for $a=2$ we 
restrict our consideration to $ 0 \leq a  <2$.  Note that in this range
$f(\xi=i, a) = \infty$, so the surface states $\langle f (a)|$ are all projectors. 
We believe that these conformal frames obey conditions $\two$ and $\three$.
Indeed $\twoa$ is  obeyed by construction since $v(i,a)=0$;
$\twob$ seems unproblematic; $\twoc$ is valid since $L^- = a K_2 + (a-1) K_4$ 
is a finite linear combination of $K_n$'s. Finally, we expect $\twod$
to hold since $\widetilde v^+$, while not completely regular,
has the same mild singularity as the butterfly.

We now ask for the values of $a$ for which  $\mathcal{L}_0$ and 
$\mathcal{L}_0^\star$ satisfy also $\one$ so that $\langle f(a)|$ 
are special projectors.  A short computation
gives:
\be
[\,\mathcal{L}_0 , \mathcal{L}_0^\star \,] = \bigl( 6a^2 -8a+4\bigr) 
\, \Bigl[ \, 2L_0 + {2a (3a-2) \over 
6a^2 -8a+4} \, (L_2 + L_{-2} ) 
+ {4(a-1)\over 6a^2 -8a+4 } ( L_4+ L_{-4} ) \,\Bigr]\,.
\ee
For the expression in brackets to be $\mathcal{L}_0 + \mathcal{L}_0^\star$
we need that the coefficients of $(L_4+ L_{-4})$ and $(L_2 + L_{-2})$ take
the right values.  For the former, comparing with (\ref{test_fam}), we get
\be
{4(a-1)\over 6a^2 -8a +4 }= (a-1) \quad\to \quad  a=1, \quad \hbox{or}\quad 
6a^2 -8a+ 4=4 ~~\to ~~ a= 0, ~  a = {4\over 3}\,.
\ee
One can readily verify that for the above three values of $a$ the coefficients
of $(L_2 + L_{-2})$ also work out.  For $a=1$ we recover the butterfly conformal
frame and $s=2$. For the other two values of $a$ we get an algebra with $s=4$.
For $a=0$ we have $\mathcal{L}_0=  L_0 - L_4$,  a familiar
projector~\cite{Gaiotto:2002kf, Schnabl:2002ff}, to be further discussed in the next subsection. 
  For $a= 4/3$ we get a new projector.

Let us now consider the conservation laws satisfied by the  states
$\langle f(a) |$.  For all values $0 \leq a <2$, we 
have 
\be
\label{ghjekjjvkj}
\langle f(a)| \mathcal{L}_{-n}=0  \,,  \quad ~n\geq -1\,.  
\ee
The above
state that $\langle f(a)|$ is the vacuum in the conformal
frame $f(\xi)$ -- these conservations hold for any surface state.
For special projectors, however, there is an additional conservation.
 We showed in \S\ref{limit_state_99} that special projectors $\langle P|$ can be obtained as the
$\gamma \to \infty$ limit of 
\be
\label{this_is_not_always_true}
\langle P| = \lim_{\gamma\to \infty}\, \langle\mathcal{I}| 
e^{-\gamma (\mathcal{L}_0 + \mathcal{L}_0^\star) } \,  \,. 
\ee
This equation implies that 
\be
\label{is_this_true}
\langle P|(\mathcal{L}_0 + \mathcal{L}_0^\star)= 0\,.
\ee 
We now demonstrate that in the family of projectors constructed
above, only the special ones satisfy this extra conservation law.  The requisite
relation can be derived by
considering the expansions 
\be
\mathcal{L}_{-n} = \oint {d\xi\over 2\pi i} 
{ (f(\xi,a))^{-n+1} \over f'(\xi,a)} \,T(\xi)
= L_{-n} + {1\over 2}\, a\, (2+n) L_{-n+2} + \ldots 
\ee
 With the help of such
relations for $\mathcal{L}_0, \mathcal{L}_{-2},$ and $\mathcal{L}_{-4}$
and (\ref{ghjekjjvkj}) one can show that 
\be
0= \langle f(a)|\Bigl(\mathcal{L}_0 + \mathcal{L}_0^\star + 
{1\over 6} a (a-1)  (3a-4) \bigl[ (2a-5)\,L_2 +  (a-3) \, L_4\bigr] + 
\ldots\Bigr) \,.
\ee
We see that, apart from $(\mathcal{L}_0 + \mathcal{L}_0^\star)$,
the additional terms shown above vanish only for the
special projectors.  One can show that for the special projectors the 
terms indicated by dots also  vanish so that the states are annihilated
by $\mathcal{L}_0 + \mathcal{L}_0^\star$.  The above, however, is sufficient to
conclude that (\ref{is_this_true}) and 
(\ref{this_is_not_always_true}) do not hold for general projectors. 
For the projectors in this family that are not special, we do not know what the state
on the r.h.s. of (\ref{this_is_not_always_true}) is.

\bigskip

As argued above, a special projector must also be annihilated
by $K$.  We want to show that this can fail to happen if
the projector is not special. It is not easy to test this claim
for the above family of states since their $L^+$ contains
a finite
number of operators and, consequently,  $K$ has
an infinite number of operators. 
To build a testable example we begin with a derivation $K$ that includes
a finite number of Virasoro operators and construct the associated $L^+$ 
and surface state.  We  take
\be
K = -3\,{\pi\over 2} \,  K_3 = -3\, {\pi\over 2}\,  (L_3 + L_{-3}) 
 \quad \to \quad  \widetilde {v^+} = -3\, {\pi\over 2}\,  (\xi^4 + {1\over \xi^2})\,,
\ee
where the constant of proportionality has been selected so that the dual vector 
corresponding to $\mathcal{L}_0 + \mathcal{L}_0^\star$ is well normalized
\be
v^+ (\xi) = -3\,  (\xi^4 + {1\over \xi^2}) 
\Bigl(\tan^{-1}\xi + \tan^{-1} ({1\over \xi}) \,\Bigr) 
= \ldots  + 2\xi  + \ldots\,.
\ee
One can check numerically that the operators $\mathcal{L}_0$ 
and  $\mathcal{L}_0^\star$ do {\em not} satisfy the algebra $\one$.
The vector $v$ corresponding to $\mathcal{L}_0$ can be read directly from
the above
\be
v(\xi) = \xi - {18\over 5}\,  \xi^3 - {18\over 7}\, \xi^5 + {2\over 3} \, \xi^7 + \ldots
\ee
By integration of $f/f' = v$ we obtain the series expansion for
the function $f$ that defines the (non-special) projector $\langle f|$:
\be
f(\xi) = \xi + {9\over 5}\,  \xi^3  + {963\over 175} \,\xi^5 +{29471\over 1575} \, \xi^7
+  \ldots
\ee
We then derive the conservation laws
\be
\begin{split}
0=& \langle{f} | \Bigl(\, \,L_{1}\, - \,\,{9\over 5}\, L_3 \, +  \ldots \,\Bigr), 
\\[0.5ex]
0=& \langle{f} | \Bigl( L_{-1} - {27\over 5}\, L_1 + {288\over 175}\, L_3 + \ldots \Bigr), 
\\[0.5ex]
0=& \langle{f} | \Bigl( L_{-3} - 9\,L_{-1} + {99\over 5}\, L_1 + {1471\over 175}\, L_3 
+ \ldots \Bigr) \,.
\end{split}
\ee
These equations imply that $0\not=  \langle{f} | K_3$, which is what we 
wanted to demonstrate. If the projector is not special $K= \widetilde L^+$ 
need not annihilate it.

\subsection{The example of  butterflies}\label{sec_butt_ex}

The butterfly state $\langle \mathcal{B} | =  \langle 0| e^{-{1\over 2} L_2} \,$
 is the projector with local coordinate $f(\xi) = \xi/\sqrt{1+ \xi^2}$. 
For the butterfly we have $v = f/f'=\xi + \xi^3$ and consequently  
\begin{equation}
{\cal L}_0 = L_0 + L_2 \qquad \hbox{and} \qquad  {\cal L}_0^\star = 
L_0 + L_{-2} \,. 
\end{equation}
In general, 
\begin{equation}
\label{fromcomm}
\mathcal{L}_n
= e^{{1\over 2} L_2}\,L_n \,
e^{-{1\over 2} L_2}\,.
\end{equation} 
By construction  we manifestly have the conservation laws:
\begin{equation}
\label{bracons}
\langle\mathcal{B } |\, \mathcal{L}_n = 0 \,, \quad \hbox{for}
\quad n \leq 1  \,.
\end{equation} 
It follows from (\ref{fromcomm}) that $\mathcal{L}_{-2k}$ with $k\geq -1$ consists
of a finite linear combination of Virasoro operators in which the highest moded 
operator is $L_2$.  Indeed, $\mathcal{L}_2 = L_2$,  $~\mathcal{L}_0 = 
L_0 + L_2$ and, quite interestingly,
\begin{equation}
\label{ummmmm}
\mathcal{L}_{-2} =  L_{-2} + 2L_0 +  L_2  = 
\mathcal{L}_0 + \mathcal{L}_0^\star \, .
\end{equation}
It follows that the butterfly $\langle\mathcal{B } |$ is annihilated by both $\mathcal{L}_0$
and $\mathcal{L}_0^\star$.
The butterfly is also annihilated 
by  $\mathcal{L}_0- \mathcal{L}_0^\star = L_2 - L_{-2} \equiv K_2$.
Equation (\ref{ummmmm}) implies the algebra $\one$ since 
\begin{equation}
[ \mathcal{L}_0 \,, \mathcal{L}_{-2} \,] = 2\mathcal{L}_{-2} 
\quad \to \quad [ \mathcal{L}_0 \,, \mathcal{L}_{0}^\star \,] = 2
\,(\mathcal{L}_0 + \mathcal{L}_0^\star ) \,. 
\end{equation}
With
$L \equiv {\textstyle {1\over 2}} \mathcal{L}_0$ and $\LL 
\equiv  {\textstyle {1\over 2}} \mathcal{L}_0^\star $
we have the canonically normalized algebra $\one$.  

\medskip
The butterfly-based interpolating
family $\langle P_\alpha|$ is defined by (\ref{palpha_states}) 
using the butterfly $L$ and~$\LL$.  The conformal map $f_\alpha(\xi)$ 
corresponding to the state $\langle P_\alpha|$ is obtained using 
the general result (\ref{sol_function}).  A computation gives  
\begin{equation}
\label{butt_fam_form}
f_\alpha (\xi) = {\xi \sqrt{1+ \alpha + \alpha\, \xi^2} \over 
1+ \alpha + (\alpha-1) \xi^2} \,.
\end{equation}
For reference we also give the $\mathcal{L}_0$ operator as a function of $\alpha$
and the surface state: 
\begin{equation}
\mathcal{L}_0 (f_\alpha) =  L_0 + {\alpha-2\over \alpha+1} \,L_2
+ {2\over (1+ \alpha)^2} \bigl( L_4 - L_6 + L_8 - L_{10} +\ldots\bigr)\,.
\end{equation}
\begin{equation}
\langle P_\alpha| = \langle 0 | e^{-A_\alpha} \,, \quad
A_\alpha = {1\over 2}\, \Bigl({\alpha-2\over \alpha+1}\Bigr)\, L_2 
+ {1\over 2} \, {1\over (1+ \alpha)^2} \, L_4 - {1\over 4}\, 
\Bigl( {2+\alpha\over (1+\alpha)^3} \Bigr)\,L_6 + \ldots\,.
\end{equation}
For $\alpha=0$ we recover the $\mathcal{L}_0$ operator and 
the surface state expression of the identity state. For $\alpha
\to \infty$ we recover the butterfly $\mathcal{L}_0$ and the 
Virasoro form
of the state.

\begin{figure}
\centerline{\hbox{\epsfig{figure=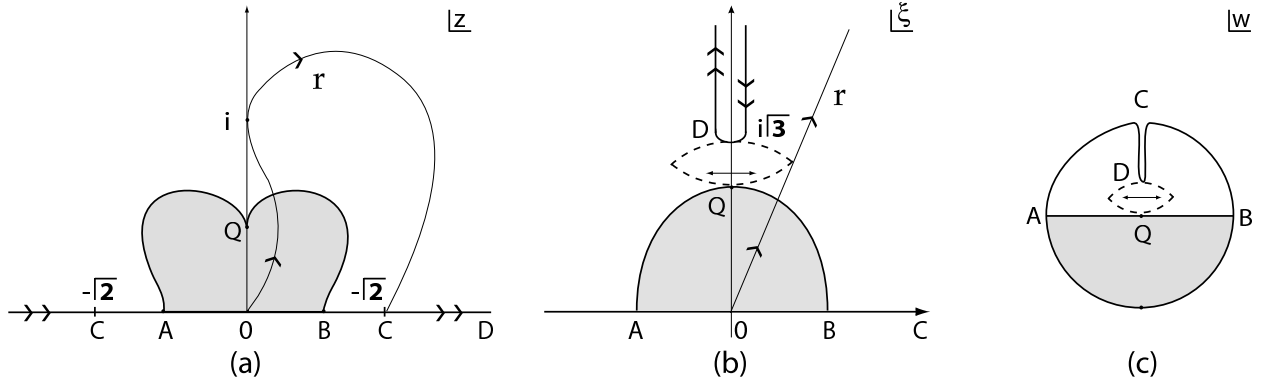, height=5.0cm}}}
\caption{The surface state  $\langle P_2|$ in the butterfly family. 
The coordinate disk in the $z$ UHP is shown in (a). The map from $\xi$
to $z$ can be extended to a full map of upper half planes if, in the
$\xi$ plane, we
cut out the region bounded by dotted lines and glue the resulting boundary
points horizontally.  
In $w$ coordinates the result is recognized as a regulated butterfly. }
\label{rz4fig}
\end{figure}

\medskip
The family of states based on the butterfly provides a natural
definition of a regulated butterfly as the state $\langle P_\alpha|$ for
$\alpha$ large.  The regulation is exact in the sense that the product of
two such regulated butterflies is a regulated butterfly and, as the regulator
is removed ($\alpha \to \infty$), we get the butterfly. The regulated butterflies
of~\cite{Gaiotto:2002kf} multiply to give regulated butterflies only approximately.
How do the states  
$\langle P_\alpha|$ look concretely?
To answer this we examine $\langle P_2|$.   For $\alpha=2$, (\ref{butt_fam_form})
gives
\be
z =f_2(\xi) = {\xi\sqrt{3+ 2\xi^2} \over 3 + \xi^2} \,.
\ee
The coordinate curve $f_2(e^{i\theta})$ is shown in Figure~\ref{rz4fig}(a).
The map $f_2(\xi)$ does not extend as a one-to-one map from the $\xi$ upper
half-plane to the $z$ upper-half plane.   In order to get a one to one map,
one must excise a small 
region bounded by dotted lines in the $\xi$-plane 
(see Figure~\ref{rz4fig}(b)). To help visualization, we show a ray $r$ that barely
touches the small region and its image in the $z$-plane. 
Since the surface cannot have a hole, there is a gluing instruction: points with equal imaginary coordinate on the dotted lines
are to be identified.  The identification is indicated in the figure with a short
$\leftrightarrow$.  Also noteworthy is that the part of the boundary beyond
$|z|=\sqrt{2}$ 
 in (a) appears as the vertical slit right above the excised region.
After mapping the $\xi$ upper-half plane to the unit $w$ disk, the picture
of the regulated butterfly is readily visualized (Figure~\ref{rz4fig}(c)).
The regulation differs from that of~\cite{Gaiotto:2002kf} only by the 
presence of an excised region.  We do not understand geometrically
why the excision makes the regulation compatible with star
multiplication.  Such understanding may follow from a
 presentation in which the excised region takes a simple shape.

\bigskip
\noindent
We examine briefly higher generalizations. Consider the conformal frames 
and associated projectors~\cite{Gaiotto:2002kf,Schnabl:2002ff}:
\begin{equation} \label{higher_butt}
z =  \xi \,\bigl( 1+(-1)^{m+1} \xi^{2m} \bigr)^{-{1\over 2m}} \,,  \qquad
\langle \mathcal{P}^{(2m)} | = 
\langle 0 |  \exp \Bigl[  (-1)^m  {1\over 2m}  L_{2m} \Bigr]\,. 
\end{equation}  
The corresponding $\mathcal{L}_0$ operators satisfy $\one$ with $s= 2m$:
\begin{equation}
\mathcal{L}_0=  L_0 + (-1)^{m+1} L_{2m} \, \quad  \to \quad [\,\mathcal{L}_0\,,
\mathcal{L}_0^\star \,] = 2m ( \mathcal{L}_0 + \mathcal{L}_0^\star)\,.
\end{equation}
Eqn.(\ref{ummmmm}) readily generalizes to
\be \label{L2m_but}
{\cal L}_{-2m} = (-)^{m+1} ( {\cal L}_0 + {\cal L}_0^\star \,) ,
\ee
and therefore  $\one$ follows from the Virasoro algebra
in the frame of the projector:
\be \label{but_expl}
[{\cal L}_0, {\cal L}_{-2m} ] = 2m {\cal L}_{-2m} \, \quad  \to \quad [\,\mathcal{L}_0\,,
\mathcal{L}_0^\star \,] = 2m ( \mathcal{L}_0 + \mathcal{L}_0^\star)\,.
\ee

The states $\langle P^m_\alpha|$ associated with these special
projectors have conformal frames  
\begin{equation}
f_\alpha^m (\xi) = {\xi \bigl[ 1+ \alpha + \alpha (-1)^{m+1} \,\xi^{2m}
\bigr]^{{1\over 2m}} \over 
\bigl[ 1+ \alpha + \alpha (-1)^{m+1} \,\xi^{2m}
\bigr]^{{1\over m}}-\xi^2 } \,.
\end{equation}
The corresponding $\mathcal{L}_0 (f_\alpha^m)$ operator takes the form
\begin{equation}
\begin{split}
\mathcal{L}_0 (f_\alpha^m) =&~~  L_0 - {2\over (1+\alpha)^{{1\over m}}}\, L_{2}
+ {2\over (1+\alpha)^{{2\over m}}}\, L_{4}  -~\ldots ~ + (-1)^{m-1}\,  
{2\over (1+\alpha)^{1-{1\over m}}}\,\, L_{2m-2}\\[0.5ex]
&+(-1)^{m+1}\, {\alpha-2\over \alpha+1}\, L_{2m} 
+ {2\over m}(-1)^m\sum_{k=1}^m {(m-k)\alpha - m \over (1+\alpha)^{1+ {k\over m}}} L_{2m+2k}
+ ~~ \ldots     
\end{split}
\end{equation}
The states $\langle \mathcal{P}^{(2m)} |$ are an infinite family of special projectors.

\sectiono{Conformal frames for 
$[{\cal L}_0, {\cal L}_0^\star] = s\, ({\cal L}_0+ {\cal L}_0^\star )$}\label{hard_work}

We have seen that condition $\one$, which states that the operators $\mathcal{L}_0$
and $\mathcal{L}_0^\star$ form the algebra 
\begin{equation}
\label{thealgebra}
[\mathcal{L}_0\,, \mathcal{L}_0^\star \,]
= s (\mathcal{L}_0+ \mathcal{L}_0^\star )\,,
\end{equation}
 is necessary for
the kinetic operator $\mathcal{L}_0$ to have simple action on
string fields of the form $f(L^+) | \mathcal{I}\rangle$.   Since $\mathcal{L}_0^\star$
is readily obtained by BPZ conjugation of $\mathcal{L}_0$, it is simply
the choice of $\mathcal{L}_0$ that determines if the 
algebra (\ref{thealgebra}) holds.  Moreover, the choice of $\mathcal{L}_0$
is the choice of a conformal map $z = f(\xi)$, as indicated in (\ref{l0_vect}).

It is the purpose
of this section to determine the class of {\it special conformal
frames}, defined as the 
functions $f(\xi)$
that result in operators $\mathcal{L}_0$
and $\mathcal{L}_0^\star$ that satisfy (\ref{thealgebra}). Note that
we do not
impose a priori conditions $\two$ and $\three$ so,
as we shall see in concrete examples, special
conformal frames need not be projectors. 
 The
set of special conformal frames is characterized by continuous
parameters for each integer $s$, but the number  of special projectors may  
be finite for each $s$.

The analysis that follows has several parts.  We derive the relevant 
constraint in \S\ref{der_cons_67}. We analyze the constraint
and relate it to the Riemann-Hilbert problem in \S\ref{sol_cns_ged}.
Solutions are presented 
in \S\ref{exp_sol_for_conf}, where
we look at examples of special projectors for $s=1$, $s=2$, and $s=3$.
In \S\ref{discuss_the_gen_status} we discuss a general pattern that seems to emerge from the examples:
the operator $\mathcal{L}_{-s}$ is always related by a ``generalized'' duality
to the operator $L^+$.  This duality makes $\one$ a consequence of the 
Virasoro commutator $[\mathcal{L}_0 , \mathcal{L}_{-s} ] = s \mathcal{L}_{-s}$.
We also construct an interesting infinite family of special projectors.

\subsection{Deriving the constraint}\label{der_cons_67}

Our strategy is to show that the algebra (\ref{thealgebra})
implies a second-order differential equation for 
$f(\xi)$. Happily, the differential equation can be integrated
twice to give a tractable condition on $f(\xi)$. 
In fact, the condition constrains the values of $f$ on the unit
circle $|\xi| =1$.

We begin the work by considering the vector fields associated
with the operators $\mathcal{L}_0$
and $\mathcal{L}_0^\star$. Equation (\ref{tv-lo_not}) furnishes the
vector $v(\xi)$ associated with $\mathcal{L}_0$:
\be
\label{first_vector}
 \quad v (\xi)  =  {1\over (\ln f (\xi))'} \,
\ee
Let us now determine the vector $v^\star(\xi)$ associated
with $\mathcal{L}_0^\star$.  Using (\ref{BPZ_vectorp}) and
letting  primes denote derivatives with respect to the argument
\begin{equation}
\label{second_vector}
v^\star(\xi) =- \xi^2\,  v(-1/\xi)= - \xi^2\, 
{f(-1/\xi) \over f'(-1/\xi)} =  {-1 \over \partial_\xi \ln f(-1/\xi)}
= {-1 \over (\ln f\circ I (\xi))' }\,,\quad \hbox{with}  \quad
I(\xi) = -{1\over \xi}\,.
\end{equation}
Given (\ref{st_comm}), we will realize the algebra (\ref{thealgebra})
if we have
\begin{equation}
[\,v (\xi) \,,  v^\star(\xi) \, ]=
-s (v(\xi) + v^\star (\xi) ) \,.
\end{equation}
This condition gives the equation
\be
v  \partial_\xi v^\star- 
v^\star \partial_\xi v  = - s ( v + v^\star) \,.
\ee
This involves first derivatives of the vectors, so second derivatives of
$f(\xi)$. A little algebra using (\ref{first_vector}) and (\ref{second_vector}) gives 
\begin{equation}
{\bigl( \ln \circ f\circ I \bigr)''\over 
\bigl( \ln \circ f\circ I \bigr)'} -
{\bigl( \ln \circ f \bigr)''\over 
\bigl( \ln \circ f \bigr)'} =  s\Bigl( \bigl( \ln \circ f \bigr)'-
\bigl( \ln \circ f \circ I \bigr)' \Bigr)\,. 
\end{equation}
After  integration and some rearrangement we find
\begin{equation}
{\bigl( \ln \circ f\circ I \bigr)'  (f \circ I)^s\over 
\bigl( \ln \circ f \bigr)' f^s } = -(-1)^s\, C_1 \,, 
\end{equation}
where $C_1$ is a constant and the sign factor has been introduced for convenience. 
This can be integrated once again to give the final equation, with
two constants of integration: 
\begin{equation}
\label{cfsa1x}
\Bigl[ f \Bigl(-{1\over \xi}\Bigr)  \Bigr]^s  + (-1)^s
 C_1  \, \bigl[f(\xi)\bigr]^s  =(-1)^s C_2 \,.
\end{equation}
Since the function $f(\xi)$ is only guaranteed to exist for $|\xi|\leq 1$, this
equation is only a condition on the values of the function 
$f(\xi)$ on the circle $|\xi| = 1$. As before,  to emphasize this point 
we  use $t= e^{i\theta}$ to denote points on the unit circle. Noting that  
$1/t^*= t$, we  write
\begin{equation}
\label{cfsa1}
\bigl[ f (-t^*)  \bigr]^s  + (-1)^s
 C_1  \, \bigl[f(t)\bigr]^s  =(-1)^s C_2 \,.
\end{equation}

 As we will see below, 
in general  $C_1$ and $C_2$ cannot be constants over the circle.
The solutions $f(\xi)$ that we find (that include the sliver and butterfly frames) 
are singular on a set of 
points on the unit disk.  These points break the circle in a set of intervals;
$C_1$ and $C_2$ need only be constants over each of
those intervals.

\subsection{Solving the constraint}\label{sol_cns_ged}

We begin our analysis of (\ref{cfsa1}) 
by imposing some 
conditions on the functions $f(\xi)$.  
Since $f(\xi)$ maps the real boundary of the  
$\xi$ half-disk to the real axis on the $z$-plane, we must have
\begin{equation}
\label{cconj}
f(\xi^*) = [f(\xi)]^* \,,
\end{equation}
where $*$ denotes  complex conjugation.  
For simplicity, we further restrict our analysis to
twist invariant  surface states,
\begin{equation}
\label{ftwistinv}
f(-\xi) = - f(\xi) \,.
\end{equation}
Equations (\ref{ftwistinv}) and (\ref{cconj}) imply that
\begin{equation}
f(-\xi^*) = - [f(\xi)]^* \,.
\end{equation}
This equation has a clear geometrical meaning: points that are reflections
about the imaginary $\xi$-axis map to points that are reflections
about the imaginary $z$-axis. We conclude that 
the coordinate curve for the map $f$ (the image under $f$ of the curve $|\xi|=1, 
\hbox{Im}(\xi) \geq 0$)   
is symmetric under reflection about the imaginary
$z$-axis. Note also that  $[f(i)]^* = f(i^*) = f(-i) = - f(i)$, so 
$f(i)$, if finite, must be purely imaginary.  

\medskip
The above conditions on $f(\xi)$  show that equation (\ref{cfsa1})
must be handled with care.  For example, letting $t \to -t$ 
 makes the left-hand side go to $(-1)^s$ times itself, so for
$s$ odd $C_2$ cannot be a constant over the whole circle. Since
$f(-t) = -f(t)$, we need only consider
 (\ref{cfsa1}) for the half-circle Re$(t)\geq 0$:  if 
$f$ is determined there it is known over the rest of the circle.
Because of the complex conjugation relation (\ref{cconj}) we can  restrict 
ourselves further to 
$t= e^{i\theta}$ with $0 \leq \theta \leq {\pi\over 2}$. 
We look for solutions in which this quarter circle is split into $N$ intervals by 
points $\theta_i$ where the function becomes singular
\begin{equation} 
\label{intervals9}
0= \theta_0 < \theta_1 < \ldots <\theta_{N-1} < \theta_N = {\pi\over 2}\,.
\end{equation}
Using (\ref{cconj}) for the first term in (\ref{cfsa1}) and cancelling
a common factor of $(-1)^s$ we obtain the conditions
\begin{equation}
\label{cfsa}
\boxed{
\begin{split}
&\phantom{~\Biggl(}\Bigl[ \bigl(f (t) \bigr)^s \Bigr]^*  + 
 C_1(k)  \, \bigl(f(t)\bigr)^s  = C_2(k) \,, \\
&\quad t = e^{i\theta} , ~~
\theta_{k-1} \leq \theta \leq \theta_k \,,  ~~k=1, \ldots, N\quad  \\
& \quad f(-t) = - f(t)\,,\qquad 
f(t^*) = (f(t))^*\,.\phantom{~\Biggl(}
\end{split}}
\end{equation}
As indicated by their argument $k$, both $C_1$ and $C_2$ can take
different (constant) values over the intervals.  We now recognize
the main condition above as a case of the general Riemann-Hilbert
problem for a disk. In this problem one looks for an analytic function
$\Phi(\xi) $ on the disk $|\xi|<1$. On the boundary $|\xi|=1$ of the disk the 
function and its complex
conjugate satisfy a relation of the form 
\be
\label{rh_problem}
(\alpha(t)+ i\beta(t))\Phi (t) + 
(\alpha(t)-i\beta(t)) (\Phi (t))^*  = \gamma (t)\,, 
\ee
for real functions
$\alpha(t), \beta(t)$ and $\gamma(t)$, with $\alpha^2 + 
\beta^2 \not=0$~\cite{muskhelishvili}.  
As we show next, over each interval
the constants $C_1$ and $C_2$ must take values consistent with the structure
of (\ref{rh_problem}).
To streamline the notation we use
\begin{equation}
F(\xi ) \equiv  \, \bigl[ f(\xi) \bigr]^s \,,
\end{equation} 
and we rewrite the main condition as
\begin{equation}
[F(t)]^* + C_1(k)  \, F(t)  = C_2(k) \,,
\end{equation}
or, more briefly, as
\begin{equation}
\label{691}
F^* + C_1 \, F  = C_2 \,.
\end{equation}
Taking the complex conjugate and dividing by $C_1^*$ we get
\begin{equation}
\label{691r}
F^* + {1\over C_1^*} \, F  = {C_2^*\over C_1^*} \,.
\end{equation}
Taking the difference of  the last two equations we find 
\begin{equation}
\Bigl( C_1 - {1\over C_1^*} \Bigr)  \, F =  C_2 - {C_2^*\over C_1^*}\,.
\end{equation}
A constant $F(t)$ is not a satisfactory solution because it implies
that $f(t)$ is unchanged as $t$ varies. 
We therefore need $C_1 C_1^*=1$, or 
\begin{equation}
C_1(k) = e^{2i\alpha_k}\,,  \quad  \alpha_k ~ \hbox{real} \,.
\end{equation}
In this case the left-hand sides of equations (\ref{691}) and (\ref{691r}) 
are identical, and the equality of the right-hand sides gives
$C_2= C_1 C_2^*$. This implies that $C_2$ is given by
\begin{equation}
C_2(k) = 2r_k\,e^{i\alpha_k}\,,
\quad  r_k ~ \hbox{real} \,.
\end{equation}
Back in (\ref{691}) we get 
\begin{equation}
\label{691xx}
F^* + e^{2i\alpha_k} \, F  = 2r_k\,e^{i\alpha_k}\,,
\end{equation}
or, equivalently,
\begin{equation}
\label{691xx22}
 e^{i\alpha_k} \, F + e^{-i\alpha_k}F^*  = 2r_k\,\,.
\end{equation}
We now see that this equation is precisely of the Riemann-Hilbert
form (\ref{rh_problem}). For our problem, the functions $\alpha,\beta$,
and $\gamma$ are piecewise constants.\footnote{In order to obtain an analytic
function that is not singular anywhere on the boundary of the disk, one must
have continuous functions $\alpha, \beta$, and $\gamma$, or more precisely,
the functions must satisfy Holder conditions~\cite{muskhelishvili}. Our functions 
$f(\xi)$ can have singularities on the boundary of the disk.}  It follows from
(\ref{691xx22}) that
\begin{equation}
\label{691xxx}
\hbox{Re}\bigl[\,F\,e^{i\alpha_k}\,\bigr] = r_k\,.
\end{equation}
We have therefore shown that
\begin{equation}
\label{ne999form}
\boxed{\phantom{\Biggl(}
\hbox{Re}\Bigl[ \bigl( f(t)\bigr)^s\,\,e^{i\alpha_k}\Bigr]
= r_k\,, \quad t = e^{i\theta} , ~~
\theta_{k-1} \leq \theta \leq \theta_k \,,  ~~k=1, \ldots, N\,.~}
\end{equation}
The condition is very simple:  over each interval the function
$u= (f(t))^s$  must lie on a straight line $\ell_k$ in the $u$-plane. 
The line is specified by the (largely) arbitrary constants $\alpha_k$ and
$r_k$.  
 The value of $\alpha_k$ is the rotation angle about $u=0$ that
makes $\ell_k$ vertical. The value of $r_k$
is the value of the real coordinate for that vertical line. A 
solution is specified by fixing some $N\geq 1$, the angles
$\theta_1, \ldots \theta_{N-1}$ that fix the intervals, and constants
$\alpha_k$ and $r_k$ for each interval.

The above prescription provides the coordinate curve that defines
the function $f(\xi)$ but does not provide the function itself.
There are two ways to obtain this function. 
In the first way, we  solve the corresponding
Riemann-Hilbert problem, which, in general, expresses the solution
in terms of fairly involved Cauchy integrals.  We will not attempt to
do so here, although we have verified that this procedure works
as expected for the case of the sliver.  In the second way,  the function
$f(\xi)$ is determined by the conformal map that takes the upper-half
$\xi$ disk to the coordinate disk.

While the prescription indicated below (\ref{ne999form}) 
to build a coordinate curve provides a solution, the solution
is sometimes formal and not always produces an $f(\xi)$ with operators
$\mathcal{L}_0$ and $\mathcal{L}_0^\star$ that satisfy (\ref{thealgebra}). 
There are conditions on the ranges
of allowed angles, certainly on the interval that contains $\xi=i$.
We leave the discussion of the matter incomplete and proceed to
illustrate with examples some large classes of solutions that we 
have checked are not formal.

\subsection{Explicit solutions for special frames}\label{exp_sol_for_conf}

We now turn to describe some concrete examples of special frames and special
projectors. We illustrate the existence
of special frames that are not special projectors by certain
deformations of the sliver.
We will not attempt to classify all special projectors. 
It appears that for a fixed $s$ positive and integer there
is a finite number of special projectors.  For $s=1$ we only
find one, the sliver.
 For $s=2$, besides the familiar
butterfly, we find an interesting new projector, the moth.  For $s=3$ 
we discuss two examples. 
All the cases that we study exhibit a common pattern: the operator ${\cal L}_{-s}$ in the frame
of the special projector plays a crucial role.

\medskip
\noindent
\subsubsection{The case $s=1$}\label{s=1subsection}

 The condition here is simply that the
coordinate curve is made of piecewise linear functions.
We only need to describe the curve to the right of the imaginary axis
because the rest of the curve is obtained by reflection.
The simplest and most familiar solution is provided by the vertical
line Re$(z) = {\pi/4}$ and Im$(z) \geq 0$.  The coordinate disk maps
to the strip $-{\pi\over 4}\leq \hbox{Re}(z)\leq {\pi\over 4}$
and Im$(z) \geq 0$.  We recognize
this as the rectangular strip of the sliver map:
\begin{equation}
f(\xi ) = \tan^{-1} (\xi)  \,, ~~\quad  (s = 1)\,. 
\end{equation}
The corresponding $\mathcal{L}_0$ operator is  
\begin{equation}
\label{slivers_lo_cal}
\mathcal{L}_0 = 
L_0 + \sum_{k=1}^\infty  {2 (-1)^{k+1}\over 4k^2-1} \, L_{2k} =
L_0 + {2\over 3} L_2 - {2\over 15} L_4 +{2\over 35} L_6 - 
\ldots~~.
\end{equation}

Another solution is provided by the horizontal line $\hbox{Im} (z) = {\pi\over 4}$.
In this case the coordinate disk maps to the strip 
$0\leq \hbox{Im}(z)\leq {\pi\over 4}$,  One can check that the mapping
function is 
\begin{equation}
\label{hor_sliv}
f(\xi ) = \tanh^{-1} (\xi)  \,, ~~\quad  (s = 1)\,. 
\end{equation}
The $\mathcal{L}_0$ operator for this map is obtained from the sliver 
$\mathcal{L}_0$  
by reversing the sign of each Virasoro operator whose mode number is
twice odd:
\begin{equation}  
\mathcal{L}_0 = 
L_0 - \sum_{k=1}^\infty  {2 \over 4k^2-1} \, L_{2k}
= L_0 - {2\over 3} L_2 - {2\over 15} L_4 -{2\over 35} L_6 - 
\ldots~~.
\end{equation}
Since $f(\xi=i)$ is finite the surface state is not a projector. The vector
$v= (\xi^2-1) \tanh^{-1} \xi$ corresponding to $\mathcal{L}_0$ does not vanish at $\xi=i$.

\begin{figure}
\centerline{\hbox{\epsfig{figure=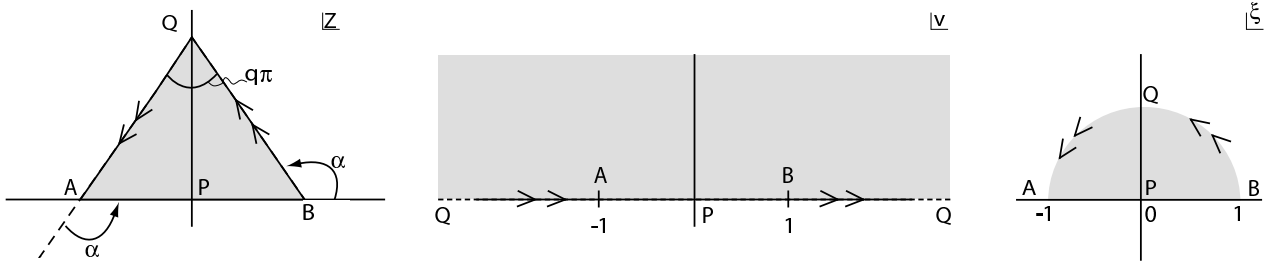, height=3.5cm}}}
\caption{Conformal maps for the $q$-deformations of the sliver.
The angle at the apex of the triangle (leftmost figure) is $\pi q$.  
The sliver is recovered for $q=0$.}
\label{charcurves}
\end{figure}

We now discuss a family of special frames that interpolate between
the sliver and the ``horizontal sliver"  (\ref{hor_sliv}).
Consider the case where the coordinate curve in the $z$ plane is the 
isosceles triangle
with vertices $A$ and $B$ on the real line and $Q$ on the imaginary axis.
The angle at $Q$  is  $q\pi$ with  $0 \leq q \leq 1$.  As we traverse the edges of 
the triangle in the counterclockwise direction the turning angles at $A$ and $B$ are
both equal to $\alpha$, where
\be
{\alpha\over \pi} = {1\over 2} (q + 1)\,.
\ee  
The map $z= f(\xi)$ is constructed in two steps.  In the first one we build 
a map from the interior of the triangle to the upper half $v$-plane with the edges
of the triangle  going to the real line and the base of the triangle mapping to
the real segment between $v=-1$ and $v=1$.  The Schwarz-Christoffel differential
equation is
\be
{dz\over dv}  = A (v-1)^{-{\alpha\over \pi}} (v+1)^{-{\alpha\over \pi}}\,,
\ee
where the magnitude of $A$ can be adjusted since it just fixes the arbitrary
scale in the $z$ plane.  We thus take $|A|$=1 and choose the  phase of $A$ such that
the equation becomes 
\be
{dz\over dv}  = { 1\over (1-v^2)^{{\alpha\over \pi}}} \,.
\ee
We thus write
\be
z= \int_0^v  {du\over (1-u^2)^{(q+1)/2}}\,.
\ee
The integral is readily expressed in terms of hypergeometric functions:
\be
z=   v\,\cdot \,{}_2\hskip-1pt F_1 \Bigl[\,\,  {1\over 2}  ,\, {1\over 2} (q+1) ,\,
 {3\over 2} ,\, v^2 \Bigr]\,.
\ee
This is the desired map from the upper-half $v$ plane to the triangle.
The map from the coordinate half-disk $\xi$ to the upper-half $v$ plane is
\be
v=  {2\xi\over 1+ \xi^2} \,.
\ee
This is, in fact, the map that defines the nothing state \cite{Gaiotto:2002kf}.  The full map we 
are looking for is simply the composition of the two maps above:
\be
z (\xi) =  {2\xi\over 1+ \xi^2}\,\cdot \,{}_2\hskip-1pt F_1 
\Bigl[\,\,  {1\over 2}  ,\, {1\over 2} (q+1) ,\,
 {3\over 2} ,\, {4\xi^2\over (1+ \xi^2)^2} \Bigr]\,.
\ee
Although complicated, series expansions are readily found:
\be
z(\xi) =   \xi  + \Bigl( -{1\over 3} + {2\over 3} q \Bigr) \xi^3 
+ \Bigl( {1\over 5} - {2\over 5} q + {2\over 5} q^2 \Bigr) \, \xi^5
+ \ldots
\ee
\be
\mathcal{L}_0 (q	) = L_0 +  \Bigl( {2\over 3} - {4\over 3} q \Bigr) L_2
+ \Bigl( -{2\over 15} - {16\over 15} q + {16\over 15} q^2 \Bigr) \, L_4 + \ldots\,,
\ee
and the surface state is 
\be
\langle \Sigma_q|  = \langle  0 |  \exp ( M) \,,
\ee
where
\be
M = -\Bigl( {1\over 3}- {2\over 3} q\Bigr)  \, L_{2}  
+ \Bigl( {1\over 30} + {4\over 15} q - {4\over 15} q^2\Bigr) L_4
+ \Bigl( -{11\over 1890} + {29\over 315} q - {76\over 315} q^2  + {152\over 945} 
q^3\Bigr) L_6+ \ldots\,.
\ee
We have tested numerically that the operators $\mathcal{L}_0(q)$ 
and $\mathcal{L}_0^\star(q)$ satisfy the algebra (\ref{thealgebra}) 
with $s=1$ in the range $q\in [0,1]$.  The $q$-deformed sliver states
$\langle \Sigma_q| $ are not projectors because $f(\xi=i)$ is finite
for $q\not=0$.  
In fact, near the midpoint one has
\be
f(\xi) = f_0 + f_1 (\xi - i)^q  + \ldots\,,
\ee
with $q$-dependent constants $f_0$ and $f_1$. 
For $q\not=0$, $f_0$ is finite, and  near the midpoint the vector field
$v(\xi)$ associated with $\mathcal{L}_0(q)$ takes the form
\be
v(\xi) = {f(\xi)\over f'(\xi)} = {f_0\over qf_1} (\xi-i)^{1-q} + \ldots
\ee
The vector $v(\xi)$ has a fractional power zero at the string midpoint.
So will the BPZ dual vector $v^\star$, the vector $v^+= v+ v^\star$ associated
with $L^+$, and the vector $\widetilde{v^+}= v^+\epsilon$ associated with $K$.
We believe that $K$ fails to annihilate the
identity. The reason is the following: since the $q$-deformed slivers are
not projectors they are not special projectors.  Thus they must fail to
satisfy at least one of the three conditions  listed in the introduction.
They satisfy condition~$\one$ and  condition~$\twoa$. The projector property
does not require $\twoc$.  So, the $q$-deformed slivers fail to satisfy
$\twob$ or $\twod$, or both.  Property $\twob$ does not strike us as
too delicate, so we feel the culprit is $\twod$ -- the failure of $K$ to
kill the identity.

For similar reasons we suspect that the coordinate curve cannot
have corners anywhere.  If it did,  $v(\xi)$ would have fractional
power zeros and those may result in a $K$ that does not kill the identity.
If this is indeed the case, we have the conclusion that the sliver
provides the unique special projector for $s=1$.

\subsubsection{The case $s=2$}

For $s=2$ we have found two special projectors: the butterfly and
a new projector, the moth. 

\begin{figure}
\centerline{\hbox{\epsfig{figure=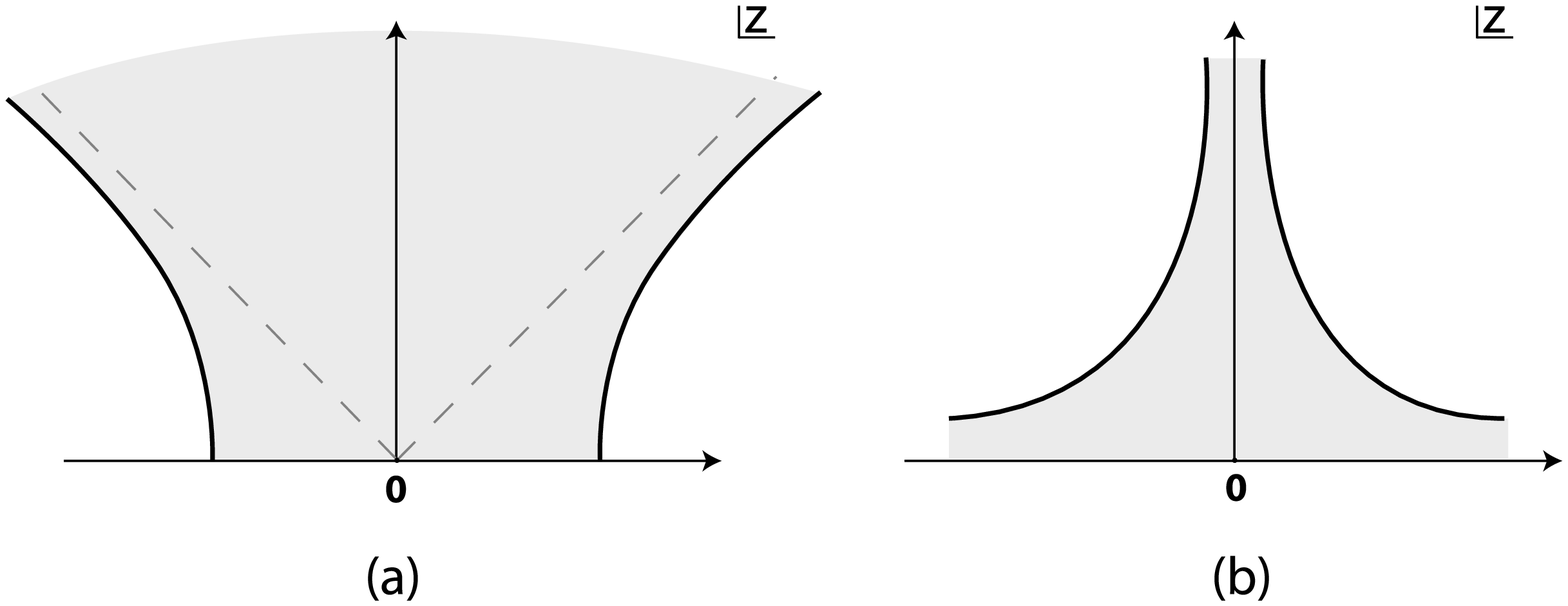, height=4.5cm}}}
\caption{Coordinate curves for $s=2$ special projectors.
(a) The coordinate curve for the butterfly. 
(b) The coordinate curve for the moth (\ref{hor_proj}).} 
\label{rz2fig}
\end{figure}

The butterfly  is a familiar example. Indeed, 
 with
\begin{equation}
f(\xi) = {\xi\over \sqrt{1+ \xi^2}}\,,
\end{equation}
we  verify that for $\xi= e^{i\theta}$ 
\begin{equation}
(f(\xi))^2 =  {e^{2i\theta} \over 1 + e^{2i\theta}}
= {e^{i\theta}\over 2 \cos \theta} = {1\over 2} + {i\over 2} \tan\theta \,. 
\end{equation}
Since this is a line (in fact vertical), the butterfly map 
satisfies  the condition (\ref{ne999form}) for $s=2$.  Note that, 
up to an irrelevant scale, the butterfly coordinate curve  is the 
square root of the sliver coordinate curve (see Figure~\ref{rz2fig}(a)).

A natural generalization of the sliver suggests itself:
\be
\label{sliver_analog}
z = f(\xi)  = \bigl[ \tan^{-1} \xi^2 \bigr]^{1/2}\,.
\ee
The square of this function, $\tan^{-1} \xi^2$, maps the circle $|\xi|=1$
to vertical lines, just like the sliver function $\tan^{-1} \xi$ does.
Thus the constraint (\ref{ne999form}) is satisfied with $s=2$.  On the
other hand, we do not get a projector since $f(\xi=i)$ is finite.

The following function, however, works all the way: 
\be
\label{hor_proj}
z = f(\xi) = \Bigl[ {1\over 2} \ln 
\Bigl( {1+ \xi^2\over 1-\xi^2} \Bigr) \Bigr]^{1/2}\, = \bigl(\tanh^{-1}\xi^2
\bigr)^{1/2}
=  \xi + 
{1\over 6} \,\xi^5 + {31\over 360}\, \xi^9 + \ldots \,.
\ee
One can readily check that $(f(\xi))^2$ maps the unit circle to a horizontal
line, thus satisfying the constraint (\ref{ne999form})  with $s=2$.
Moreover $f(\xi=i)$ is infinite.  So, we have a special projector 
(see Figure~\ref{rz2fig}(b)), which we shall call the moth. 
A quick computation gives 
an $\mathcal{L}_0$ reminiscent of the sliver's (\ref{slivers_lo_cal}): 
\be
\mathcal{L}_0  = L_0 + \sum_{k=1}^\infty  {2 (-1)^k\over 4k^2-1} \, L_{4k} \,.
\ee
Numerically tests indicate that, as expected,
 the algebra (\ref{thealgebra}) is satisfied. 
With $v$ denoting the vector  
associated with $\mathcal{L}_0$ one finds that:
\be
\label{moth's_v+}
v^+ = {1\over 2} ( v + v^\star) = {1-\xi^4\over 2\xi} \,\,\bigl(
\tanh^{-1} (\xi^2) - \tanh^{-1} (1/\xi^2) \bigr) \,.
\ee

We can use (\ref{sol_function}) to calculate the
maps $f_\alpha(\xi)$ that define the $\langle P_\alpha |$ states
of the  moth family.  Including a convenient overall normalization
constant, we find
\be \label{falphas2}
f_\alpha (\xi) =\frac{ \sqrt{1+\alpha} }{2}
\Bigl[ \, \Bigl( \frac{1+ \xi^2}{1-\xi^2} \Bigr)^{\frac{2}{1+ \alpha}} 
  -1\, \Bigr]^{1/2} \,.
\ee
This family of states was  encountered before in \cite{Gaiotto:2002kf}, 
where it appeared  as 
the most general family of (twist invariant)
 surface states annihilated by 
the derivation 
$K_2 = L_2- L_{-2}$\footnote{Comparing with eqn.(9.10) of \cite{Gaiotto:2002kf}, we find that
the parameter $\mu$ in (9.10) is related to the parameter 
$\alpha$ in (\ref{falphas2}) as
$\mu = -1/(1 + \alpha)$. In the present
context $\alpha \geq 0$, hence $-1 \leq \mu \leq 0$. 
In the context of~\cite{Gaiotto:2002kf},
$\mu>0$ is legitimate, and one recognizes {\it e.g.} the 
butterfly state for $\mu =1/2$ and the nothing state for $\mu = 1$.}. 
Interestingly, a quick calculation shows that  ${\cal L}_{-2} = - K_2$. The Virasoro
algebra in the moth frame gives $[{\cal L}_0, K_2] =  2 K_2$, whose
BPZ conjugate is
$[{\cal L}_0^\star, K_2] = -2 K_2$. It follows that $[L^+\,,K_2] =0$, 
which explains why the whole family, constructed
as an exponential of $L^+$ acting on the identity,  is killed by $K_2$. 
The states $| P_\alpha \rangle$ are also annihilated by
 a second derivation, the operator $K = \widetilde L^+$, which is an infinite linear
 combination of $K_{2n+1}$ operators and thus
clearly different from $K_2$. Note also
 that  $ [K_2, K]= [K_2, L^+]\widetilde{\phantom{A}} = 0$.
   For $\alpha=1$
 we find  $f_1(\xi) = \xi/\sqrt{1-\xi^2}$, which
 is the butterfly map 
with the ``wrong'' sign inside the square
 root. The corresponding state is
 \be
 | P_1 \rangle = e^{\frac{1}{2} L_{-2}} \, | 0 \rangle \,.
 \ee
 This wrong-sign butterfly has been studied in \cite{Schnabl:2002ff}.
 According to \cite{Schnabl:2002ff} (eqn.(57)),
 the product of two wrong-sign butterflies
 is the surface state associated with the function (\ref{falphas2})
 with $\alpha=2$, in nice agreement with   
$|P_1 \rangle* |P_1\rangle = |P_2\rangle$.  The multiplication of an 
infinite number of wrong-sign butterfly states $|P_1\rangle$ gives the 
moth projector
$| P_\infty\rangle$ described by (\ref{hor_proj}).

\subsubsection{The case $s=3$}

For $s=3$ we construct explicitly two special projectors. 

\begin{figure}
\centerline{\hbox{\epsfig{figure=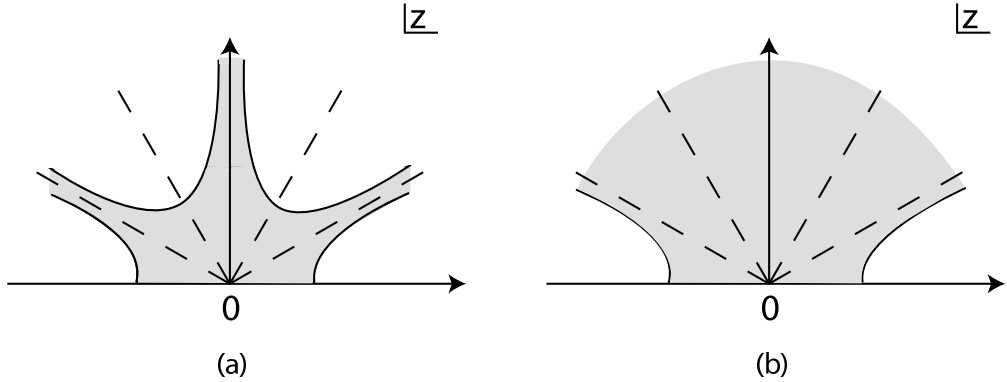, height=5.0cm}}}
\caption{Coordinate curves for $s=3$ special projectors.
(a) The coordinate curve for the (\ref{sliver_analog_s3}) 
projector. 
(b) The coordinate curve for the projector (\ref{s3_second}). 
}
\label{rz3fig}
\end{figure}

The generalization (\ref{sliver_analog}) of the sliver, that did not
work for $s=2$, works for $s=3$: 
\be
\label{sliver_analog_s3}
z = f(\xi)  = \bigl[ \tan^{-1} \xi^3 \bigr]^{1/3} = \xi - {1\over 9} \,\xi^7 + {22\over 
405} \,\xi^{13} + \ldots \,.
\ee
Indeed, condition (\ref{ne999form}) is satisfied with $s=3$
and $f(\xi=i)= \infty$.  In fact $f(\xi= e^{i\pi/6}) = \infty$. This 
example corresponds to a realization of (\ref{ne999form}) 
with $N=2$ with $\theta_1 = \pi/6$. 
The coordinate curve for this projector is shown in Figure~\ref{rz3fig}(a).
With $v$ denoting the vector 
associated with $\mathcal{L}_0$ one finds that:
\be
\label{proj_s=3's_v+}
v^+ = {1\over 3} ( v + v^\star) =  {1+\xi^6\over 3\xi^2} \,\,\bigl(
\tan^{-1} (\xi^3) + \tan^{-1} (1/\xi^3) \bigr) \,.
\ee
A quick calculation shows that
${\cal L}_{-3} =K_3 = L_3 + L_{-3}$.  
We can repeat the argument given for the moth to show that
$K_3$ commutes with $L^+$ and thus annihilates the whole
family $P_\alpha$ based on this projector. 
The family will also be killed by $K = \widetilde {L^+} \neq K_3$.

This suggests looking for a projector for which $K$ {\it is} 
proportional to ${\cal L}_{-3}$.
Such projector is a different kind of generalization of the sliver,
which satisfies  $ \mathcal{L}_{-1}\sim K$.
Moreover, $K\sim K_1$, so the vector field $(1+ \xi^2)$ 
associated with the sliver $\mathcal{L}_{-1}$ vanishes
for $\xi=\pm i$.  For the $s=3$ projector we take 
$\mathcal{L}_{-3} = K_3 + a K_1$ and adjust the constant
$a$ so that the vector associated with $\mathcal{L}_{-3}$ 
has zeroes only at $\xi= \pm i$.  This gives
$a=3$, so we get 
\be
\label{ghjkgu3}
\mathcal{L}_{-3}= K_3 + 3 K_1  \,~~ \quad  \to ~~\quad  v_{\mathcal{L}_{-3}}= 
{(1+ \xi^2)^3\over \xi^2} \,.
\ee
We also have
\be
\mathcal{L}_{-3} = \oint {d\xi\over 2\pi i}  \,{3\over (f^3)'} \,T(\xi) 
\quad \to \quad v_{\mathcal{L}_{-3}}= {3\over (f^3)'}\,. ~~~~~~
\ee
The last two equations give a differential equation
for $f$ that is readily integrated:
\be
\label{s3_second}
f(\xi) = (3/8)^{1/3}\,\Bigl[ \tan^{-1} \xi + {\xi^3-\xi\over (1+ \xi^2)^2} \Bigr]^{1/3}
= \xi- {3\over 5} \,\xi^3+ {87\over 175} \,\xi^5 + \ldots \,.
\ee
We now verify that $K$ is indeed proportional to $\mathcal{L}_{-3}$.  We first use
$f$ to calculate the vector $v$ associated with $\mathcal{L}_0$ and then form 
\be
v^+(\xi)= {1\over 3} (v + v^\star) = 
{ (1+ \xi^2)^3\over 8\xi^2} \Bigl( \tan^{-1} \xi + \tan^{-1} (1/\xi) \Bigr)\,,
\ee
so that 
\be
v_K = \widetilde{v^+}(\xi) = v^+(\xi) \,\epsilon(\xi) = {\pi\over 16}
\,{(1+ \xi^2)^3\over \xi^2} = {\pi\over 16} \,v_{\mathcal{L}_{-3}}\, 
\quad \to \quad  K = {\pi\over 16} \,\mathcal{L}_{-3}\,,
\ee
as desired.  
The relation $K \sim\mathcal{L}_{-3}$  explains why
 (\ref{thealgebra}) holds:  the Virasoro algebra commutator
$[\mathcal{L}_0, K] = 3 K$ upon dualization gives 
$[\mathcal{L}_0, L^+] = 3 L^+$.  This is equivalent
to (\ref{thealgebra}) with $s=3$.  The coordinate curve for the special projector
(\ref{s3_second}) is shown in figure Figure~\ref{rz3fig}(b).  
Note that the coordinate curve is a sub-curve
of that for the projector (\ref{sliver_analog_s3}), shown in Figure~\ref{rz3fig}(a). 
The coordinate disks differ, of course.

\medskip  
Other $s=3$ special projectors are likely to exist. 
For example, a projector whose coordinate curve (in the region Re$\,z>0$)
is the cubic root of a horizontal
line may 
 exist, according to (\ref{ne999form}).  
It would be the $s=3$ analog of (\ref{hor_proj}).

\subsection{Generalized duality}\label{discuss_the_gen_status}

In all the examples of special projectors that we
have considered the operator ${\cal L}_{-s}$
is interesting in some way.
We have seen that for 
each even $s$ there is a (higher) butterfly
projector (\ref{higher_butt}) for which ${\cal L}_{-s} \sim L^+$.
The algebra $\one$ then follows from the 
Virasoro commutator $[\mathcal{L}_0 , \mathcal{L}_{-s} ] 
= s \mathcal{L}_{-s}$. For each odd $s$
there is a ``dual''  construction
where ${\cal L}_{-s}\sim K= \widetilde {L^+}$. 
The $s=1$ and $s=3$ constructions give the sliver
and  the projector (\ref{s3_second}), respectively.
This construction generalizes
to all odd $s$, as we will discuss at the end of the present
section. The algebra $\one$ now follows from 
$[\mathcal{L}_0 , \mathcal{L}_{-s} ] 
= s \mathcal{L}_{-s}$ and duality.

The two remaining examples, the  $s=2$ moth (\ref{hor_proj}) and the $s=3$ projector (\ref{sliver_analog_s3}),
follow a somewhat different pattern. In these cases ${\cal L}_{-s} $
 is very simple -- it is proportional, respectively, to the derivations
 $K_2$ and $K_3$ -- but  has no apparent connection
 with the fundamental objects  $L^+$ and $K$.
  To establish this connection we  generalize the notion of  duality.  

 We introduce the function $\epsilon_\alpha (t)$, with $0 \leq \alpha < \pi$,
 defined for $t = e^{i \theta}$ by
 \be
 \epsilon_\alpha(e^{i \theta}) = 
\begin{cases}
 \phantom{-}1 \,,  &\text{if $\, ~ \alpha -\pi  < \theta \leq~ \alpha$~;} \\[0.5ex]
-1\,,  &\text{if $\, ~~\;~\alpha~ <~ \theta \leq ~\alpha + \pi$.}
\end{cases} 
 \ee
 In other words, 
we bisect the unit circle 
 at an  angle $\alpha$ and define $\epsilon_\alpha$ to take
 values $\pm 1$ in the two halves. The function
 $\epsilon (t)$ that we have used so far is $\epsilon_{\pi/2}(t)$.
Two functions on a circle will be said to be dual to each other if one
is equal to the other times the product of a finite number of $\epsilon_\alpha$ 
functions, with different values of $\alpha$.  
Clearly, squaring a product of $\epsilon_\alpha$'s gives the
constant function~$1$ on the circle.

Consider now the function that appears multiplicatively
in the $v^+$ vector (\ref{moth's_v+}) of the moth. We claim
that
\be
\label{moth's_v+++}
\tanh^{-1} (t^2) - \tanh^{-1} (1/t^2)  = 
-i\,\frac{\pi}{2}\,  \epsilon_{\pi/2}(t)\,  \epsilon_0 (t)\,.
\ee
Indeed, for any $t= e^{i\theta}$ on the circle the left-hand side is equal
to $\pm i\pi/2$. The signs alternate over angular intervals of ninety degrees,
with value $+i\pi/2$ for $\theta\in [0, \pi/2]$. The product on the right-hand
side reproduces this behavior. Recalling, additionally, that 
${\cal L}_{-2} = -K_2$, we have
\be
v_{{\cal L}_{-2} } (t)  =  {1-\xi^4\over \xi} \,.
\ee
It now follows from (\ref{moth's_v+}) and the last two equations that for the moth
\be
\label{gen_dual_moth}
 v^+(t) =-i\,\frac{\pi}{4}\,  \epsilon_{\pi/2}(t)\,  \epsilon_0(t)
 \; v_{{\cal L}_{-2} } (t)  \, ,
\ee
showing that after all, $v_{{\cal L}_{-2} }$ and $v^+$, as well as
$\mathcal{L}_{-2}$ and $L^+$, are related by duality.
We can now dualize the $\mathcal{L}_{-2}$ in the commutator 
$[{\cal L}_0, {\cal L}_{-2}] = 2 {\cal L}_{-2}$ to obtain
the relation  $[{\cal L}_0, L^+] = 2  L^+$. This dualization
is allowed following the logic of (\ref{duality_commutator}),
which demands that the product of the underlying vectors vanishes
at the discontinuities of the dualizing function. This holds, since
the vector corresponding to $\mathcal{L}_0$  vanishes for $\xi = \pm i$
and $\pm 1$.

Dualizing (\ref{gen_dual_moth}) by further multiplication by $\epsilon_{\pi/2}(t)$
we get 
\be \label{great}
v_K(t)= i\,\frac{\pi}{4}\, \epsilon_{0} (t) \, v_{K_2}  (t)  \, .
\ee
This means that  $K$ is proportional to the dual of $K_2$, with duality flipping 
the sign of the vector on the upper half circle $\Im t > 0$.
The vector $v_{K_2}$ vanishes at the
discontinuities   $t = \pm 1$ of $\epsilon_0(t)$,
so we can conclude that $[K, K_2]$  is dual to
the commutator $[K, K]$ and must therefore vanish. 

The $s=3$ projector (\ref{sliver_analog_s3}) can be understood similarly. 
The vector $v^+$ given in (\ref{proj_s=3's_v+}) contains the multiplicative
factor $(\tan^{-1} (t^3) + \tan^{-1} (1/t^3))$ that is equal to $\pm \pi/2$
over six alternating intervals of the unit circle, with plus sign for
$\theta \in [-\pi/6, \pi/6]$. This time one finds
the interesting relation 
\be
v^+= \frac{\pi}{6}\,  \epsilon_{\pi/2} \, \epsilon_{\pi/6}\, 
\epsilon_{5 \pi/6}  \;v_{{\cal L}_{-3}}  \, .
\ee
The vector $v_{{\cal L}_{-3}} = v_{K_3}= \xi^4+1/\xi^2$ vanishes at 
the discontinuities of the dualizing function
so, once again, $\one$ follows by 
dualization of the $[{\cal L}_0, {\cal L}_{-3}]$ commutator.

In summary, for all special projectors known  so far, 
the operator ${\cal L}_{-s}$ is related to $L^+$ (and to $K$)
by a generalized duality. This duality relation, 
 which might be generic,  ``explains" why these conformal frames are special.

\medskip
Finally, we would like to briefly describe an
interesting infinite family of special projectors, which contains
as special cases several examples that we have already discussed.
For every integer $s$ we look for the conformal frame $f$ in which
the vector field associated with ${\cal L}_{-s}$ is 
 \be \label{Lsansatz}
 v_{{\cal L}_{-s}} \equiv \frac{s}{(f^s)'}= \frac{(1 + \xi^2)^s}{\xi^{s-1}} \,.
 \ee 
 Note that $v_{{\cal L}_{-s}}$ is BPZ even (odd) for $s$ even (odd).
 Integrating (\ref{Lsansatz}), we find
 \be \label{fLs}
 f(\xi) = \xi  \left(  {}_2F_1 \left[ \frac{s}{2}, s\,; 1 +\frac{s}{2}\,;
  -  \xi^2\right]\right)^{1/s} \,.
 \ee
These conformal frames  are projectors for all real $s \geq 1$:
 $f_s(\pm i) = \infty$.

 For $s=1$ the hypergeometric function simplifies
and we recover the sliver. For $s=3$ we recover
the projector (\ref{s3_second}), 
 as we should given (\ref{ghjkgu3}).
 Using a standard hypergeometric identity we can rewrite (\ref{fLs}) as
 \be
 f (\xi) = \frac{\xi}{\sqrt{1+ \xi^2}}  \,
 \left(  {}_2F_1 \left[ \frac{s}{2}, 1-\frac{s}{2}\,;
 1 +\frac{s}{2}\,;  \frac{\xi^2}{\xi^2+1}\right]\right)^{1/s} \,.
 \ee
 This presentation makes it manifest that for $s$ even the hypergeometric
 function truncates to a finite polynomial of its argument.
 For $s=2$ we recognize the butterfly map. For $s=4$
 we recover the special projector (\ref{a_family}) with $a=4/3$. 
Curiously,  for $s=-1$ we find the map of the identity.  
  We claim that for each even $s$, the operator ${\cal L}_{-s}$  is proportional
to $L^+$, while for each odd $s$ it is proportional to $K$.
We simply quote the result, obtained using hypergeometric
identities,
\be
 \frac{\Gamma(s/2+1) \Gamma(s/2)}{\Gamma(s+1)}\,\, {\cal L}_{-s}\,=
\begin{cases}
L^+\quad {\rm for}\; s\; {\rm even} \, ,\\
K\phantom{L}\quad {\rm for}\; s\; {\rm odd}\,.
\end{cases}
 \ee
This result implies that the algebra $\one$ holds and the projectors
are special. It would be interesting to
investigate if (\ref{fLs}) defines  special projectors even for
non-integer $s$. It seems clear that we have just begun to understand the rich algebraic structure
of special projectors.

\sectiono{Concluding Remarks}

Since a  summary of our results was given in
the introductory section, we limit ourselves here to point out
some questions that remain open.

At a technical level, it would be interesting to have
a complete classification of special projectors. Perhaps the ideas
of generalized duality explained in \S\ref{discuss_the_gen_status} will turn out
to be useful. 
Explicit
forms for the function $f(\xi)$ that defines the projectors
are in general missing. We have not determined if
there are special projectors for non-integer~$s$.
A complete analysis will require understanding what are the
conditions on the vector $v$
associated with $\mathcal{L}_0$ that ensure that properties
$\twoc$ and $\twod$ hold.  
It also became clear in
our work that the framework of conservation laws in the
operator formalism requires generalization to deal
with vector fields that, having certain singularities on
the unit circle, do not define analytic functions over
the rest of the complex plane.  

It remains somewhat surprising that there is
a notion of a special projector.  For a special projector
one finds a  
family of states, built using a rather
simple prescription, that interpolate from the identity to
the projector.  One wonders if a related, perhaps
more
complicated construction, exists for arbitrary projectors.
We have noted that for arbitrary projectors the corresponding
$|P_\alpha\rangle$
states multiply as expected but $|P_\infty\rangle$ is not
the projector one starts with.

In order to make the techniques discussed here
applicable to the physical ghost-number one equation
of motion one may generalize the abelian
algebra $\mathcal{A}_f$ to include the action of ghost
oscillators.  This should suffice to construct the string
field corresponding to the tachyon vacuum.  An additional
extension of $\mathcal{A}_f$ to include matter oscillators
seems necessary to produce solutions that describe $D$-brane
solitons, Wilson lines, and the time-dependent decay of
D-branes.

In the process of extending the results to the ghost and matter
sectors we will be able to find out if there is something
truly special about the sliver that allowed Schnabl to find a solution,
or if all special projectors are on the same footing.
We think this question is an important one, since it can
eventually help simplify the solution and understand better
its universal features.
Such understanding is likely to
point out ways to obtain new and perhaps unexpected
solutions of open string field theory.

\vspace{1cm}

{\bf \large Acknowledgments}

 \medskip

We would like to thank D.~Gaiotto, J.~Goldstone, and M.~Headrick
for helpful conversations. We are grateful to Y.~Okawa, 
 M.~Schnabl, and A.~Sen  for critical reading of the manuscript
and useful suggestions. 
The work of LR is  supported in part
 by the National Science Foundation Grant No. PHY-0354776. 
  Any opinions, findings, and
conclusions or recommendations expressed in this material are
those of the authors and do not necessarily reflect the views of
the National Science Foundation.
The work of BZ is supported in part
by the U.S. DOE  grant DE-FC02-94ER40818.

\medskip

\appendix

\sectiono{Lie Algebra and Lie Group relations}

\noindent
We consider the nonabelian Lie algebra with two generators
\begin{equation}
[\, L \,, \, \LL \, ] =  L+ \LL \,,
\end{equation}
and the corresponding adjoint representation
\begin{equation}
L = \begin{pmatrix}  0&1 \cr 0 & 1 \end{pmatrix}\,, \quad
\LL = \begin{pmatrix}  -1&0 \cr -1 & 0 \end{pmatrix}\,.
\end{equation}
As matrices we have the following relations
\begin{equation}
L L = L \,, \quad  \LL\LL = - \LL  \,, \quad
L \LL = \LL \,, \quad \LL L = - L \,.
\end{equation}
Note also the relation
\begin{equation}
(L+ \LL)^2 = 0 \,.
\end{equation}
In this representation group elements are given by
\begin{equation}
e^{\alpha L + \beta \LL} =  1 +  {e^{\alpha-\beta} -1\over \alpha-\beta}\,
\, (\alpha L + \beta \LL) \,.
\end{equation}
\noindent
Particular useful cases are
\begin{equation}
x^L =  1 + (x-1) L \,, \quad  y^{\LL} = 1 + 
\Bigl( 1 - {1\over y}\Bigr) \,\LL \,.
\end{equation}

\medskip
\noindent
The CBH formula for this group can be derived by comparing
group elements in the adjoint representation.  We find that 
\begin{equation}
e^{\alpha L}  \, e^{\beta \LL} = e^{ \beta' \LL}\, e^{\alpha' L}  \,
=  e^{\tilde\alpha L + \tilde\beta \LL} 
\end{equation}
determines 
$(\tilde\alpha,\tilde\beta)$ in terms of $(\alpha, \beta)$ or in
terms of
$(\alpha', \beta')$:
\begin{equation}
\begin{split}
\tilde \alpha~ & = ~{\alpha-\beta\over e^{-\beta} - e^{-\alpha}} 
\,(1- e^{-\alpha})
~ = ~{\alpha'-\beta'\over e^{\alpha'} - e^{\beta'}} 
\,(e^{\alpha'}-1)\,, \\
\tilde \beta~ & = ~{\alpha-\beta\over e^{-\beta} - e^{-\alpha}} 
\, (1- e^{-\beta})
~= ~{\alpha'-\beta'\over e^{\alpha'} - e^{\beta'}} 
\, (e^{\beta'}-1)\,.
\end{split}
\end{equation}

\noindent
We also have the following inverse relations:
\begin{equation}
\begin{split}
e^{\alpha} &=  \Bigl( 1 - {\tilde \alpha\over \tilde \beta}\Bigr)^{-1} \,
\Bigl[ 1 -  {\tilde \alpha\over \tilde \beta} \, e^{\tilde \alpha- \tilde 
\beta} \Bigr]\,~~, \quad
e^{\beta} =  \Bigl( 1 - {\tilde \alpha\over \tilde \beta}\Bigr)^{-1} \,
\Bigl[ e^{\tilde \beta-\tilde \alpha}  -  
{\tilde \alpha\over \tilde \beta} \, \Bigr]\,, \\
e^{\alpha'} &=  \Bigl( 1 - {\tilde \alpha\over \tilde \beta}\Bigr) \,
~~\Bigl[ 1 -  {\tilde \alpha\over \tilde \beta} \, e^{\tilde \beta- \tilde 
\alpha} \Bigr]^{-1}\,, \quad
e^{\beta'} =  \Bigl( 1 - {\tilde \alpha\over \tilde \beta}\Bigr) \,
~~\Bigl[ e^{\tilde \alpha-\tilde \beta}  -  
{\tilde \alpha\over \tilde \beta} \, \Bigr]^{-1}\,.
\end{split}
\end{equation}
We could also give the relations $(\alpha, \beta) \leftrightarrow
(\alpha',\beta')$.  But it is more useful to write them as Schnabl, 
who gives the first one of the following:
\begin{equation}
\begin{split}
x^L \, y^{\LL}  & = \Bigl( {y\over x+ y - xy}  \Bigr)^{\LL} 
\,\Bigl( {x\over x+ y - xy} \Bigr)^{L} \,, \\[1.0ex]
y^{\LL} \, x^{L}  &= \Bigl( {x+ y - 1\over y}\Bigr)^{L} ~~
\,\Bigl( {x+ y - 1\over x}\Bigr)^{\LL}\,.
\end{split} 
\end{equation}

\noindent
Useful corollaries are 
\begin{equation}
\label{diffls}
 x^{L - \LL}  = \Bigl( {2\over 1+ x^2}\Bigr)^{\LL} 
\Bigl( {2x^2\over 1+ x^2}\Bigr)^L =  
\Bigl( { 1+ x^2\over 2}\Bigr)^L 
\Bigl( {1+ x^2\over 2x^2}\Bigr)^{\LL}
\end{equation} 
as well as the fairly redundant but helpful identities
\begin{equation}
\label{simpeid}
\begin{split}
x^{\LL}   x^L &= ~  e^{\bigl( 1 - {1\over x}\bigr) (L + \LL) }
~=~ \Bigl( 2 - {1\over x} \Bigr)^{L} \,
\Bigl( 2- {1\over x} \Bigr)^{\LL} \,,
\\[0.8ex]
x^L  \, x^{\LL} &= ~  e^{( x-1) (L + \LL) }
~~=~ 
\Bigl( {1\over 2-x}\Bigr)^{\LL} \,\,\, \Bigl( {1\over 2-x}\Bigr)^{L} \,,
\\[0.9ex]
e^{x ( L + \LL)} & =~ 
\Bigl( {1\over 1-x}\Bigr)^{\LL}\, \Bigl( {1\over 1-x}\Bigr)^{L} ~
= ~  (x+1)^{L} \, (x+1)^{\LL} \,.
\end{split}
\end{equation}

\begingroup\raggedright

\providecommand{\href}[2]{#2}

\end{document}